\documentclass[aps,prd,a4paper,preprintnumbers,amsmath,amssymb,superscriptaddress,twocolumn,floatfix,showpacs]{revtex4}
\usepackage{amssymb}
\usepackage{float,graphicx}
\newcommand{\Oo}{{\cal O}}
\newcommand{\dd}{{\rm d}}

\newcommand{\GeV}{\,{\rm GeV}}

\newcommand{\kpc}{\,{\rm kpc}}
\newcommand{\pc}{\,{\rm pc}}

\newcommand{\eV}{\,{\rm eV}}

\newcommand{\Cin}{{\rm Cin}}
\newcommand{\Si}{{\rm Si}}

\newcommand{\etal}{\emph{et al.}\,}
\newcommand{\Pphi}{\mathcal{P}_{\gamma \leftrightarrow \phi}}
\newcommand{\Pbar}{\bar{\mathcal{P}}_{\gamma \leftrightarrow \phi}}
\newcommand{\be}{\begin{eqnarray}}
\newcommand{\ee}{\end{eqnarray}}
\newcommand{\ba}{\left( \begin{array}{ccc}}
\newcommand{\ea} {\end{array} \right)}
\newcommand{\bv}{\left( \begin{array}{c}}
\newcommand{\ev} {\end{array} \right)}
\newcommand{\hatb}[1]{\hat{\mathbf{#1}}}
\begin{document}
\title{The Effect of a Chameleon Scalar Field on the Cosmic Microwave Background}
\author{Anne-Christine Davis}
\email{a.c.davis@damtp.cam.ac.uk}
\affiliation{Department of Applied Mathematics and Theoretical Physics,
Centre for Mathematical Sciences,  Cambridge CB3 0WA, United Kingdom}
\author{Camilla A.O. Schelpe}
\email{C.A.O.Schelpe@damtp.cam.ac.uk}
\affiliation{Department of Applied Mathematics and Theoretical Physics,
Centre for Mathematical Sciences,  Cambridge CB3 0WA, United Kingdom}
\author{Douglas J. Shaw}
\email{D.Shaw@qmul.ac.uk}
\affiliation{Queen Mary University of London, Astronomy Unit, Mile End Road, London E1 4NS, United Kingdom}
\date{\today}
\begin{abstract}
We show that a direct coupling between a chameleon-like scalar field and photons can give rise to a modified Sunyaev--Zel'dovich (SZ) effect in the Cosmic Microwave Background
(CMB). The coupling induces a mixing between chameleon
particles and the CMB photons when they pass through the magnetic field of a galaxy cluster.  Both the intensity and the polarization of the radiation are modified. The degree of
 modification depends strongly on the properties of the galaxy
cluster such as magnetic field strength and electron number density. Existing SZ measurements of the Coma cluster enable us to place constraints on the photon-chameleon coupling. 
The constrained conversion probability in the cluster is 
$ {\mathcal{P}}_{\rm Coma}(204\,{\rm GHz}) < 6.2 \times 10^{-5}$ at $95\%$ confidence, corresponding to an upper bound on the coupling strength of  $g_{\rm eff}^{\rm (cell)} < 2.2\times 10^{-8}\,{\rm GeV}^{-1}$ or 
$g_{\rm eff}^{\rm (Kolmo)} < (7.2-32.5) \times 10^{-10}\,{\rm GeV}^{-1}$, depending on the model that is assumed for the cluster magnetic field structure. We predict the 
radial profile of the chameleonic CMB intensity decrement. We find that the chameleon effect extends further towards the edges of the cluster than the thermal SZ effect. Thus we 
might see a discrepancy between the X-ray emission data and the observed SZ intensity decrement. We further predict the expected change to the CMB polarization arising from the existence of a chameleon-like scalar field. These predictions could be verified or constrained by future CMB experiments.
\end{abstract}
\maketitle
\section{Introduction}
Measurements of the current cosmic acceleration of the universe 
have led to theories of an unknown energy component with negative
pressure in the Universe. This dark energy is generally modelled by
a scalar field rolling down a flat potential. On Solar-system
scales this scalar field should appear effectively massless. The origin
of the dark energy scalar field is still highly speculative; some
suggestions have involved the moduli fields of string theory.
However, if a near-massless scalar field existed on Earth it should
have been detected as a fifth force unless the coupling to normal
matter was unnaturally suppressed. Khoury and Weltman
\cite{Khoury04} introduced the chameleon scalar field which has gravitational strength or greater coupling to normal matter but
a mass which depends on the density of surrounding matter. On cosmological
scales the chameleon successfully acts as a low-mass dark energy candidate,
while in the laboratory its large mass allows it to evade detection. 

An extension to the original chameleon model is to include an interaction
term between the chameleon and electromagnetic (EM) field of the form
$\frac{\phi}{M_{\rm eff}}F_{\mu\nu}F^{\mu\nu}$, first suggested by Brax
\emph{et al.\,}\cite{Brax07}. This leads to interconversion between
chameleons and photons in the presence of a magnetic field, which
alters the intensity and polarization of any radiation entering the
magnetic region. A coupling between the chameleon and EM fields gives
rise to observational effects that could potentially be detected either
astrophysically or in the laboratory \cite{Brax07,BurrageSN,Burrage08}.
These predictions have lead to constraints on the mass parameter $M_{F}$
describing the coupling strength to the EM field: $M_F\gtrsim2\times10^{6}\,\mathrm{GeV}$
\cite{Brax07} and $M_F\gtrsim1.1\times10^{9}\,\mathrm{GeV}$ \cite{Burrage08}.  

It was noted in Ref. \cite{Burrage08} that in low density backgrounds, such as the interstellar and intra-cluster mediums, the 
mixing between the chameleon field and photons is indistinguishable from the mixing of any very light scalar field with a linear coupling to $F_{\mu \nu}F^{\mu \nu}$ or a very 
light pseudo-scalar with a linear coupling to $\epsilon_{\mu \nu\rho \sigma}F^{\mu \nu}F^{\rho \sigma}$.  Such fields are collectively referred to as axion-like particles (ALPs).
 Standard ALPs have the same mass and photon-coupling everywhere. The best constraints on their parameters come from limits on their production in relatively high density regions
  such as stellar cores or in supernova 1987A \cite{ALPrev}.  Chameleon scalar fields represent one example of a subclass of ALPs that we term chameleonic ALPs. The defining 
  feature of chameleonic ALPs is that the astrophysical constraints from high density regions are evaded or exponentially weakened.  This may be because as the ambient density 
  increases the field mass rises as in the chameleon 
  model, or because the effective photon coupling weakens  \cite{Jaeckel07, Burrage08}.  This latter possibility is realized in the Olive--Pospelov model for a density-dependent fine-structure constant \cite{Olive08}.  The results we derive in this paper technically apply to all ALPs, however the parameter range for which they apply is strongly ruled out for all but chameleonic ALPs. The model for  chameleon-like theories is described in detail in subsection \ref{sub:The-Basic-Chameleon} below. 

In this paper we consider the propagation of the Cosmic Microwave
Background (CMB) radiation through the magnetic field of a galaxy
cluster. We predict the modification to the CMB intensity and polarization
in the direction of galaxy clusters arising from mixing with a chameleonic ALP.

One of the significant foreground effects acting on the CMB as it
passes through a galaxy cluster, which has been extensively verified
by experiment, is the Sunyaev--Zel'dovich (SZ) effect. This
arises from scattering of CMB photons off electrons in the galaxy
cluster atmosphere, causing a change to the CMB intensity in the direction
of the cluster. Measurements of the SZ effect can be compared to predictions
of the CMB intensity decrement arising from the chameleon, and thus
constrain the chameleon model parameters. 

This paper is organized as follows: in \S \ref{sec:Optics},
the conversion between photons and a general scalar ALP in a magnetic field
is analysed, and note how it depends on the structure of the magnetic field. In \S \ref{sec:ClusterMagneticField}, we briefly review the observational evidence for magnetic fields
 in the intra-cluster medium (ICM) and detail their typical properties. In \S \ref{sec:SZ}, we describe the standard SZ effect and show how this is modified when photons are converted to (chameleonic) ALPs in the intra-cluster medium. We note that the frequency and electron density dependence of the chameleonic SZ effect differs from that of the thermal and other standard contributions to the SZ effect.  In \S \ref{sec:Application}, we exploit this difference to constrain the probability of photon-scalar conversion in the ICM.  Under certain reasonable assumptions about the magnetic field structure, we convert these constraints to bounds on the effective photon to scalar coupling. In \S \ref{sec:Modification-Polarisation} we discuss the modification to the CMB polarization from chameleon-mixing. We summarize our results in \S \ref{sec:Conclusions}.  

\subsection{Chameleon-like Theories}\label{sub:The-Basic-Chameleon}
Chameleon-like-theories, which include the standard chameleon model introduced by Khoury and Weltman \cite{Khoury04} as well as the Olive--Pospelov, varying $\alpha_{\rm em}$ model \cite{Olive08}, are essentially generalized scalar-tensor theories and as such are described by the following action:
 \be
\mathcal{S}&=&\int\mathrm{d}^{4}x\sqrt{-g}\left(\frac{M_{Pl}^{2}}{2}\mathcal{R}-\frac{1}{2}g^{\mu\nu}\partial_{\mu}\phi\partial_{\nu}\phi\right. \label{action} \\
&&\left.-V(\phi)-\frac{1}{4}B_{F}(\phi/M)F_{\mu\nu}F^{\mu\nu}\right) \nonumber \\ && +\mathcal{S}_{\mathrm{matter}}\left(\psi^{(i)},B^{2}_{i}(\phi/M)g_{\mu\nu}\right)\,,\nonumber
\ee
where the $\psi_{i}$ are the matter fields and $\mathcal{S}_{\rm matter}$ is the matter action (excluding the kinetic term of electromagnetism). The $B_{i}(\phi/M)$ determine the
 coupling of the scalar, $\phi$, to different matter species; $B_{F}(\phi/M)$ determines the photon-scalar coupling. For simplicity we assume a universal matter coupling 
 $B_{i}(\phi/M) = B_{\rm m}(\phi/M)$, although we do not require that $B_{\rm m}=B_{F}$. Varying this action with respect to $\phi$ gives the scalar field equation:
\be
\square \phi = \frac{\partial V_{\rm eff}}{\partial \phi}(\phi; F^2, T_{\rm m}), \label{fieldEq}
\ee
where 
\begin{equation}
V_{\rm eff}(\phi; F^2, T_{\rm m}) \equiv V(\phi) + \frac{B_{F}(\phi)}{4} F^2 +
B_{\rm m}(\phi) T_{\rm m}. \nonumber
\end{equation}
Here $T_{\rm m} = \rho_{\rm m}-3P_{\rm m}$, is the trace of the energy-momentum tensor and corresponds to the physical, measured density or pressure. For this derivation we must 
remember that the particle masses are constant and independent of $\phi$ in the conformal Jordan frame with metric $g_{\mu\nu}^{(i)}=B^{2}_{i}g_{\mu\nu}$, while the length scales
 are measured in the Einstein conformal frame. Thus $\rho_{\rm m}=B_{i}^{3}\rho_{\mathrm{Jordan}}$.
 
Whether or not a given scalar-tensor theory is chameleon-like is determined by the form of its self-interaction potential, $V(\phi)$, and its coupling functions, 
$B_{\rm m}(\phi/M)$ and $B_{F}(\phi/M)$.  A simple and very general prescription is that the magnitude of small perturbations in $\ln B_{F}(\phi/M)$ and $\ln B_{i}(\phi/M)$ about 
their values at the minimum of $V_{\rm eff}(\phi)$, and sourced by some small perturbation in $T_{\rm m}$ or $F^2$, decreases as $T_{\rm m}$ becomes large and positive,  i.e. as the ambient density increases.  
This implies that for any combination of $i,j = F,m$,
\be
H_{ij}(F^2, T_{\rm m}) = \left \vert\frac{(\ln B_{i})_{,\phi}(\phi_{\rm min}/M)(\ln B_{j})_{,\phi}(\phi_{\rm min}/M) }{V_{,\phi\phi}^{\rm eff}(\phi_{\rm min};F^2,T_{\rm m})}\right\vert \label{Hcond}
\ee
is a decreasing function of $T_{\rm m}$, where $\phi_{\rm min}$ is given by $V_{,\phi}^{\rm eff}(\phi_{\rm min}; F^2, T_{\rm m}) = 0$. We must also require that $\phi_{\rm min}$ exists and that $V_{,\phi\phi}^{\rm eff}(\phi_{\rm min};F^2,T_{\rm m}) >0$ for stability. 

\subsubsection{The Chameleon Model}
\begin{figure}[htb!]
\begin{centering}
\includegraphics[clip,width=7.5cm]{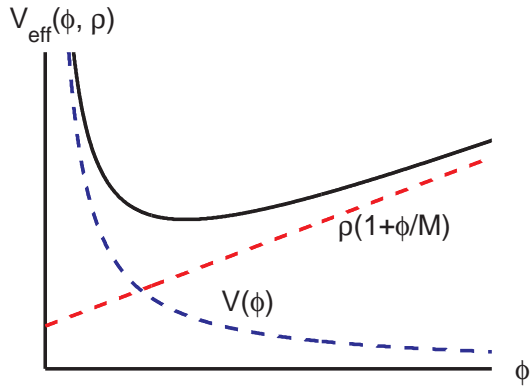}
\par\end{centering}

\caption{\label{fig:chameleon-potential}The chameleon effective potential,
$V_{\mathrm{eff}}$, is the sum of the scalar potential, $V(\phi)$,
and a density-dependent term \cite{Khoury04}.} 

\end{figure}

In a standard chameleon theory, we assume that $\phi/M \ll 1$ and approximate $B_{F} \approx 1 +
\phi/M_{\rm eff}$ and $B_{\rm m}^{2}  \approx 1 + 2\phi/M$, where this defines $M$ and $M_{\rm eff}$. We expect $M_{\rm eff} \sim \Oo(M)$ but this does not necessarily have to be
 the case.  Assuming $T_{\rm m}/M \gg \vert F^2\vert/4M_{\rm eff}$, the condition on $H_{ij}$ is then equivalent to
$$
V_{,\phi \phi\phi} < 0.
$$
This implies that the $\phi$ field equations will be non-linear in $\phi$. Additionally the existence of $\phi_{\rm min}$ and stability implies
$$
V_{,\phi}<0, \qquad V_{,\phi \phi}>0.
$$
A typical choice of potential can always be described (in the limit of $\phi \gg \Lambda$) by
\be
V(\phi)\approx \Lambda_{0} + \frac{\Lambda^{n+4}}{n\phi^{n}}, \label{Veqn}
\ee
where for reasons of naturalness $\Lambda \sim O(\Lambda_{0})$. For the chameleon field to be a suitable candidate for dark energy,
\be
\Lambda_{0} = (2.4\pm  0.3)\times 10^{-3}\,{\rm eV}.\nonumber
\ee
A corollary of Eq. (\ref{Hcond}) in chameleon theories is that the chameleon field is heavier in dense environments than it is in sparse ones. From Eq. (\ref{fieldEq}), we can see
 that the mass of small perturbations, $m_{\phi}$, in the chameleon field is simply  given by $m_{\phi}^2 = V^{\rm eff}_{,\phi \phi}$.  The chameleon field, $\phi$, rolls along 
 the effective potential, $V_{\rm eff}$, and the shape of $V_{\rm eff}$ depends on the ambient density through $T_{\rm m}$. For baryonic matter under normal conditions, $T_{\rm m}
  \approx \rho_{\rm m}$, and so when, as is often the case, $\rho_{\rm m} \gg \vert F^2 \vert$ and $\phi/M \ll 1$ we have,
\be
V_{\rm eff} \approx V(\phi) + \left(1+\frac{\phi}{M}\right)\rho_{\rm m}.\nonumber
\ee
A plot of this effective potential, for when $V(\phi)$ is given by Eq. (\ref{Veqn}) and $n=1$, is shown in Fig. \ref{fig:chameleon-potential}.
 From this we can see that the value of $\phi$ at the minimum will
decrease as the density of the surrounding matter, $\rho_{\rm m}$, increases,
while the curvature of the effective potential at the minimum will
increase. It is clear from this that the mass of the chameleon is greater in relatively high density environments such as laboratories on Earth, while in space the chameleon field
 is much lighter. It is this property that allows the chameleon field to have a strong coupling to matter, yet be light enough in space to produce interesting phenomena, such as non-negligible mixing with photons, whilst being heavy enough in the laboratory to avoid being ruled out by constraints on fifth-forces.  A detailed discussion of the allowed chameleon parameters is given in Ref. \cite{MotaShaw}.

The best direct upper-bounds on the matter coupling $1/M$ come simply from the requirement that the chameleon field makes a negligible contribution to standard quantum amplitudes,
 roughly $1/M \lesssim 10^{-4} \,{\rm GeV}^{-1}$ \cite{MotaShaw, BraxLight}. Stronger constraints have been derived on the photon coupling $g_{\rm eff} = 1/M_{\rm eff}$. Laser-based 
 laboratory searches for general ALPs, such as PVLAS \cite{PVLAS}, and chameleons, such as GammeV \cite{GammeV}, give $g_{\rm eff} \lesssim 10^{-6}\,{\rm GeV}^{-1}$.  Constraints 
 on the production of starlight polarization from the mixing of photons with chameleon-like particles in the galactic magnetic field currently provide the strongest upper-bound of
  $g_{\rm eff} < 9 \times 10^{-10}\,{\rm GeV}^{-1}$ \cite{Burrage08}.  With such couplings and for a standard choice of potential such as Eq. (\ref{Veqn}) with $n \sim O(1)$, $m_{\phi} \ll 10^{-14}\,{\rm eV}$ in the galactic and intra-cluster media.

\subsubsection{The Olive--Pospelov Model}
In any theory with the action of Eq. (\ref{action}), the effective fine-structure constant, $\alpha_{\rm em}$, is proportional to $1/B_{F}(\phi/M)$.  Observations of QSO spectra 
suggest that the $\alpha_{\rm em}$ in dust clouds roughly $10\,{\rm Gyrs}$ ago differs by a few parts in $10^{6}$ from its value today as measured in the laboratory \cite{Webb}.
  This could be explained by either a time variation \cite{VaryingAlpha} in $\alpha_{\rm em}$, or a density dependence \cite{Olive08} since the clouds are much sparser than the
  laboratory environment.  In standard chameleon theories the variation in $\alpha_{\rm em}$ is constrained to be well below the level that can be detected, and so cannot explain
  this observation.  Olive and Pospelov suggested a chameleon-like model in which a $\Oo(10^{-6})$ change in $\alpha_{\rm em}$ (or any other constant of nature) from dense to very
  sparse environments is feasible.  This model has,
\begin{eqnarray}
B_{F} &=& 1 +
\frac{\xi_{F}}{2}\left(\frac{\phi-\phi_{\rm m}}{M}\right)^2,  \nonumber \\ 
B_{\rm m} &=& 1 +
\frac{1}{2}\left(\frac{\phi-\phi_{\rm m}}{M}\right)^2, \nonumber
\end{eqnarray}
where $\xi_{F}$ is an $\Oo(1)$ constant. Whilst one may consider
more complicated choices for the potential, a simple choice satisfying 
all the required conditions is $V(\phi) = {\rm const} + m^2
\phi^2/2$. In low density environments such as the interstellar or
intra-cluster media $\phi_{\rm min} \approx 0$, while in high density environments such as the
laboratory $\phi_{\rm min}\approx \phi_{\rm m}$. The total fractional change in $\alpha_{\rm em}$ from high to low density environments in this model is then 
$\delta \alpha/\alpha \simeq \xi_{F}\phi_{\rm m}^{2}/2M^2$.

The effective photon coupling in low density environments is
\begin{eqnarray}
g_{\rm eff} &=& 1/M_{\rm eff} \equiv (\ln B_{F})_{,\phi}(\phi_{\rm min}/M) \nonumber \\ 
 &\simeq& B_{F,\phi}(\phi_{\rm min}/M) \approx \frac{10^{-3}\xi_{\rm F}^{1/2}}{M} \left\vert \frac{\delta
 \alpha}{10^{-6}\alpha} \right\vert^{1/2}. \nonumber
\end{eqnarray}
In high density backgrounds $g_{\rm eff}$ is much smaller. In
low density backgrounds, the mass of the Olive--Pospelov scalar,
$m_{\phi}\simeq m^{2}$ and the constraints found
in Ref. \cite{Burrage08} imply $m_{\phi} \ll 10^{-14}\eV$.  For the
Olive--Pospelov model to satisfy all current experimental constraints,
$1/M$ must be neither too large nor too small. Olive and Pospelov gave
the lower-bound $1/M \gg 10^{-9}\,{\rm GeV}^{-1}$, which for
$\delta\alpha/\alpha \sim O(10^{-6})$ implies $g_{\rm eff} \gg 10^{-12}\,{\rm GeV}^{-1}$. Laboratory tests such as GammeV and PVLAS are insensitive to OP scalar fields, because at their relatively high operating densities the scalar-photon coupling is greatly suppressed. The best upper-bound on $g_{\rm eff}$ comes, as with standard chameleon fields, from limits on the production of starlight polarization in the galaxy. Thus if the OP model is to explain a $10^{-6}$ change in $\alpha_{\rm em}$ as suggested by observations of QSO spectra, we must have $10^{-12}\,{\rm GeV}^{-1} \ll g_{\rm eff} < 9 \times 10^{-10}\,{\rm GeV}$.

\section{Optics with a scalar Axion-Like Particle}\label{sec:Optics}
In the presence of an axion-like particle (ALP), the properties of electromagnetic radiation propagating through a magnetic field are altered as a result of mixing between the 
ALP and photons.  Over the past 25 years or so, there has been a great deal of work on this subject; for example Refs. \cite{Silkvie84,   Raffelt88, Harai92, Burrage08} or see 
Ref. \cite{ALPrev} for a recent review. 

In this section, we present the equations which describe the mixing of a chameleon-like particle with the photon field in the presence of a magnetic field.  We also present the 
solution of these equations, in the limit where the mixing is weak,  for two different models of the magnetic field structure in the ICM.  

Varying the action in Eq. (\ref{action}) with respect to both $\phi$ and $A_{\mu}$, we find Eq. (\ref{fieldEq}) and
\begin{eqnarray}
\nabla_{\mu} \left[B_{F}(\phi/M) F^{\mu \nu}\right] &=& J^{\nu}, \nonumber
\end{eqnarray}
where  $J^{\mu}$ is the background electromagnetic 4-current; $\nabla_{\mu}J^{\mu} = 0$.  

We are concerned with the propagation of light in sparse
astrophysical backgrounds containing a magnetic field
$\mathbf{B}(x^{\mu})$. The background values of the chameleon and
photon fields are assumed to be slowly varying over length and
time scales of $\Oo(1/\omega)$ where $\omega$ is the proper frequency
of the electromagnetic radiation being considered. We define the
perturbations in the photon and chameleon fields about their
background values to be,  $a_{\mu} = A_{\mu}-\bar{A}_{\mu}$ and
$\varphi = \phi - \bar{\phi}$, respectively. When electromagnetic
radiation propagates through a plasma with electron number density
$n_{\rm e}$, an effective photon mass-squared of $\omega_{\rm  pl}^2 =
4\pi \alpha_{\rm em} n_{e}/m_{e}$ is induced; $\omega_{\rm pl}$ is the
plasma frequency. Following
Ref. \cite{Burrage08} in ignoring terms that are second order in the
perturbations $\varphi$ and $a_{\mu}$, and assuming $\omega \gg H$ where $H$ is the Hubble parameter, we find
\begin{eqnarray}
-\ddot{\mathbf{a}} + \nabla^2 \mathbf{a} &\simeq & \frac{\boldsymbol{\nabla} \varphi \times \mathbf{B}}{M_{\rm eff}} + \omega_{\rm pl}^2 \mathbf{a},  \nonumber \\
-\ddot{\varphi} + \nabla^2 \varphi &\simeq & \frac{\mathbf{B} \cdot (\boldsymbol{\nabla} \times \mathbf{a})}{M_{\rm eff}} + m_{\phi}^2 \varphi, \nonumber 
\end{eqnarray}
where $1/M_{\rm eff} \equiv  B_{F,\phi}(\bar{\phi})$ and $m_{\phi}^2 \equiv
V_{{\rm eff},\phi\phi}(\bar{\phi}, T_{\rm m}, \bar{A}_{\mu})$. 

We consider a photon field propagating in the $\hatb{z}$ direction, so
that in an orthonormal Cartesian basis $(\hatb{x}, \hatb{y},
\hatb{z})$ we have $\mathbf{a} = (\gamma_x, \gamma_y, 0)^{\rm T}$. The
above equations of motion for the chameleon and photon polarization
states can then be written in matrix form: 
\be
\left[ -\partial_{t}^{2} -\partial_{z}^{2} - \ba \omega_{\rm pl}^{2} &
  0 & -\frac{B_{y}}{M_{\rm eff}}\partial_{z} \\ 0 & \omega_{\rm pl}^{2}
  & \frac{B_{x}}{M_{\rm eff}}\partial_{z}  \\ \frac{B_{y}}{M_{\rm
      eff}}\partial_{z} & -\frac{B_{x}}{M_{\rm eff}}\partial_{z} &
  m_{\phi}^{2} \ea \right] \bv \gamma_{x} \\ \gamma_{y} \\ \varphi \ev
= 0. \nonumber
\ee
Following a similar procedure to that in Ref. \cite{Raffelt88}, we
assume the fields vary slowly over time and that the refractive index
is close to unity which requires $m_{\rm \phi}^2/2\omega^2$, $\omega_{\rm
  pl}^2/2\omega^2$ and $\vert B\vert /2\omega M_{\rm eff}$ all
$\ll 1$. Defining $\gamma_{i} = \tilde{\gamma}_{i}(z)
e^{i\omega(z-t)+i \beta(z)}$ and $\varphi = \tilde{\varphi}
e^{i\omega(z-t)+i \beta(z)}$ where $\beta_{,z} = -\omega_{\rm
  pl}^2(z)/2\omega$,  we approximate $-\partial_{t}^{2}\approx
\omega^{2}$ and $\omega^2+\partial_{z}^{2}\approx 2\omega (\omega
+i\partial_{z})$. In addition to this, we define the state vector
\be
\mathbf{u} = \bv \tilde{\gamma}_{x}(z) \\ \tilde{\gamma}_{y}(z) \\ e^{2i\Delta(z)}\tilde{\varphi}(z)\ev. \nonumber
\ee
where $$\Delta(z) =  \int_{0}^{z} \frac{m_{\rm eff}^2(x)}{4\omega} \dd x, $$
and $m_{\rm eff}^2 = m_{\phi}^2(z) - \omega_{\rm pl}^2(z)$, which
simplifies the above mixing field equations for $\mathbf{a}$ and
$\varphi$ to 
\be
\mathbf{u}_{,z} = \frac{\mathcal{B}(z)}{2M_{\rm eff}} \mathbf{u}, \label{stateEqn}
\ee
where
\be
\mathcal{B}(z) = \ba 0 & 0 & -B_{y}e^{-2i\Delta} \\ 0 & 0  &B_{x}e^{-2i\Delta} \\ B_{y}e^{2i\Delta} & -B_{x}e^{2i\Delta} & 0 \ea. \nonumber
\ee
To solve this system of equations we expand, $\mathbf{u} =
\mathbf{u}_{0}+\mathbf{u}_{1}+\mathbf{u}_{2}+\ldots$, such that
\be
{\mathbf{u}}_{i+1}' = \frac{\mathcal{B}(z)}{2M_{\rm
    eff}}\mathbf{u}_{i}\,;\;{\mathbf{u}}_{0}' =0\,, \nonumber
\ee
and neglect the higher order terms from mixing. We define 
$$\mathcal{M}_{1} = \int_{0}^{z} \frac{\mathcal{B}(x)}{2M_{\rm eff}} \dd x. $$
We say that mixing is weak when
\be
{\rm tr}\left( \mathcal{M}_{1}^{\dagger}(z)\mathcal{M}_{1}(z) \right) \ll 1. \nonumber
\ee
Thus Eq. (\ref{stateEqn}) can be solved in the limit of weak-mixing: 
\be
\mathbf{u}(z) \simeq \left[\mathbb{I}+\mathcal{M}_{1}(z) + \mathcal{M}_{2}(z)\right] \mathbf{u}(0), \nonumber
\ee
where $ \mathcal{M}_{2} =
\int_{0}^{z}\mathcal{M}_{1}^{\prime}(x)\mathcal{M}_{1}(x) \dd x$, which
can be written
\be
\mathbf{u}(z) \simeq \left[\mathbb{I}+\mathcal{M}_{1}(z) + \frac{1}{2}\left(\mathcal{M}_{C}(z)+\mathcal{M}_{1}^2(z)\right)\right] \mathbf{u}(0), \nonumber
\ee
where explicitly
\be
\mathcal{M}_{1} &=& \ba 0 & 0 & -A_{y}^{\ast} \\ 0 & 0 & A_x^{\ast} \\ A_y & -A_x & 0 \ea, \nonumber \\
\mathcal{M}_{C} &=& \ba -C_{yy} & -C^{\ast}_{xy} & 0 \\ C_{xy} & -C_{xx} & 0 \\  0 & 0 & C_{xx}+C_{yy} \ea, \nonumber
\ee
and 
\be
A_{i} &=& \int_{0}^{z} \dd x \frac{B_{i}(x)e^{2i\Delta(x)}}{2M_{\rm eff}}, \nonumber \\ 
C_{ij} &=& \int_{0}^{z} \dd x\, \left( A_{i}^{\ast \prime}(x)A_{j}(x) - A_{i}^{\ast}(x)A_{j}^{\prime}(x)\right). \nonumber
\ee  
We note that $\mathcal{M}_{n}^{\dagger} = -\mathcal{M}_{n}$. 

The polarization state of radiation is described by its Stokes
parameters: intensity $I_{\gamma}(z) = \vert\gamma_{x}(z)\vert^2 +
\vert \gamma_{y}(z)\vert^2$, linear polarization $Q(z) =
\vert\gamma_{x}(z)\vert^2 - \vert \gamma_{y}(z)\vert^2$ and $U(z) =
2{\rm Re} \left(\gamma_{x}^{\ast}(z)\gamma_{y}(z)\right)$, and
circular polarization $V(z) = 2{\rm Im}
\left(\gamma_{x}^{\ast}(z)\gamma_{y}(z)\right)$. Assuming there is no initial chameleon flux, $I_{\phi} = \vert \varphi\vert^2 = 0$, then to leading order the final photon intensity is given by
\be
I_{\gamma}(z) &=& I_{\gamma}(0)\left(1-\Pphi(z)\right) \nonumber \\ &&+ Q(0) \mathcal{Q}_{\rm q}(z) + U(0) \mathcal{Q}_{\rm u}(z) + V(0) \mathcal{Q}_{\rm v}(z), \nonumber
\ee
where we have defined,
\be
\Pphi(z) &=& \frac{1}{2}\left(\vert A_{x}(z)\vert^2 + \vert A_{y}(z)\vert^2\right), \nonumber \\
\mathcal{Q}_{\rm q}(z) &=& \frac{1}{2}\left(\vert A_{x}(z)\vert^2 - \vert A_{y}(z)\vert^2\right), \nonumber \\
\mathcal{Q}_{\rm u}(z) &=& {\rm Re}(A_{x}^{\ast}A_{y}), \nonumber \\
\mathcal{Q}_{\rm v}(z) &=& {\rm Im}(A_{x}^{\ast}A_{y}).\nonumber
\ee
In this article we consider mixing with the CMB, the intrinsic polarization of which is small,
i.e. $\sqrt{Q^2(0)+U^2(0)+V^2(0)}/I_{\gamma}(0) \ll 1$.  It follows
that to leading order the modification to the Stokes parameters of
the CMB radiation, from the effect of photon-ALP mixing, is
\be
\Delta\mathbf{S}(z) = \Delta \bv I_{\gamma}(z) \\ Q(z) \\ U(z) \\ V(z) \ev &\approx& \bv -\Pphi(z) \\ \mathcal{Q}_{\rm q}(z) \\ \mathcal{Q}_{\rm u}(z) \\
 \mathcal{Q}_{\rm v}(z) \ev I_{\gamma}(0). \nonumber
\ee
To evaluate $\Pphi$ and the $\mathcal{Q}_{i}$ we need a model for the spatial variation of the magnetic field and effective mass.   Typically $n_{\rm e} \sim 10^{-3} \text{--} 
10^{-2}\,{\rm cm}^{-3}$ in galaxies and galaxy clusters and so $\omega_{\rm pl} \sim \Oo(10^{-12})\eV$.  For the chameleon-like-theories detailed in 
\S \ref{sub:The-Basic-Chameleon}, it follows that $\omega_{\rm pl}^2 \gg m_{\phi}^2$. In what follows we assume that the scalar field is very light so that 
$m_{\rm eff}^2 \approx -\omega_{\rm pl}^2 \propto -n_{\rm e}$. Spatial variations in $m_{\rm eff}^2$ are then entirely due to spatial variations in $n_{\rm e}$.  

The magnetic fields of galaxies and galaxy clusters have been measured to have a regular component $\mathbf{B}_{\rm reg}(z)$, with a typical length scale of variation similar to 
the size of the object, and a turbulent component $\delta\mathbf{B}(z)$ which undergoes $\Oo(1)$ variations and reversals on much smaller scales.  It is common, both in the 
literature concerning photon-ALP conversion in galaxy and cluster magnetic fields and in that concerning the measurement of those fields, to use a simple \emph{cell model} to 
describe $\delta\mathbf{B}$.  In this cell model one defines $L_{\rm coh}$ to be the length scale over which the random magnetic field is coherent, i.e. the typical scale 
over which field reversals of $\delta\mathbf{B}$ occur.  A given path of length $L$ is then divided into $N$ magnetic domains, $N=L/L_{\rm coh}$. The magnetic field strength and 
electron density are assumed to be constant along the path, but the random magnetic field component, $\delta\mathbf{B}_{x}$, is taken to have a different random orientation in 
each of the $N$ magnetic domains.  Although the cell model for $\delta\mathbf{B}$ is common in the literature and reveals most of the key features associated with astrophysical 
photon-ALP mixing, it is not accurate especially when $\vert \Delta(z=L)\vert \gg 1$.  Additionally, as was first noted in Ref. \cite{Ensslin03}, in this model we do not have  $\nabla \cdot \delta\mathbf{B} = 0$ as we should.  A more realistic model for $\mathbf{B}$ is to assume a spectrum of fluctuations running from some very large scale (e.g. the scale of the galaxy or cluster) down to some very small scale ($\ll L_{\rm coh}$).  We describe these fluctuations by a correlation function $R_{\rm B}(\mathbf{x}) = \left\langle \delta \mathbf{B}(\mathbf{y})\delta \mathbf{B}(\mathbf{x}+\mathbf{y})\right\rangle$ and the associated  power spectrum $P_{\rm B}(k)$; here the angled brackets indicate the expectation of the quantity inside them.  We can, and should, also allow for a similar spectrum of fluctuations in the electron number density with some power spectrum $P_{\rm N}(k)$.  We now evaluate $\Pphi$ and the $\mathcal{Q}_{i}$ in both this \emph{power spectrum model} and the cell model.

\subsection{The Cell Model}\label{CellModel}
In the cell model, we take $m_{\rm eff}^2 = {\rm const}$ and
$\mathbf{B} = \mathbf{B}_{\rm reg} + \delta\mathbf{B}$ where
$\mathbf{B}_{\rm reg} \approx {\rm const}$ over the path length
$L$. For $\delta \mathbf{B}$ we take $(\delta\mathbf{B})_{x} = B_{\rm
  rand} \cos \theta_{n}$ and $(\delta\mathbf{B})_{y} =  B_{\rm
  rand}\sin \theta_{n}$ for $(n-1)L_{\rm coh} < z < nL_{\rm coh}$
 where the $\theta_{n}$ are independent random variables with identical distributions $U(0, 2\pi]$. 
We then have, over a path length $L$, 
\be
A_{i} &=&  (\mathbf{A}_{\rm reg})_{i} + (\delta \mathbf{A})_{i}, \nonumber \\ 
\mathbf{A}_{\rm reg} &=& \frac{2\mathbf{B}_{\rm reg} \omega e^{i\bar{\Delta}}}{M_{\rm eff}m_{\rm eff}^2}\sin\bar{\Delta}, \nonumber  \\
\delta \mathbf{A} &=& \frac{2B_{\rm rand} \omega e^{\frac{i\bar{\Delta}}{N}}}{M_{\rm eff}m_{\rm eff}^2}\sin\left(\frac{\bar{\Delta}}{N}\right) \sum_{r = 0}^{N-1} \hatb{n}^{(r)} e^{\frac{2i \bar{\Delta} r}{N}}, \nonumber
\ee
where $\bar{\Delta} = m_{\rm eff}^2 L/4\omega$ and $\hatb{n}^{(n)} = (\cos \theta_n,\sin \theta_n,0)^{\rm T}$. It follows that the expected values of $\Pphi$ and the $\mathcal{Q}_{i}$ are,
\be
\Pbar &\equiv& \left \langle \Pphi \right\rangle = \frac{1}{2}\left(\frac{2 B_{\rm reg} \omega }{M_{\rm eff}m_{\rm eff}^2}\right)^2\sin^2\bar{\Delta} \label{Pphibox} \\&&+ \frac{N}{2}\left(\frac{2B_{\rm rand} \omega}{M_{\rm eff}m_{\rm eff}^2}\right)^2\sin^2\left(\frac{\bar{\Delta}}{N}\right), \nonumber  \\
\bar{\mathcal{Q}}_{\rm q} &=& \frac{\cos 2\theta_{\rm reg}}{2}\left(\frac{2 B_{\rm reg} \omega }{M_{\rm eff}m_{\rm eff}^2}\right)^2\sin^2\bar{\Delta}, \nonumber \\
\bar{\mathcal{Q}}_{\rm u} &=& \frac{\sin 2\theta_{\rm reg}}{2}\left(\frac{2 B_{\rm reg} \omega }{M_{\rm eff}m_{\rm eff}^2}\right)^2\sin^2\bar{\Delta}, \nonumber\\
\bar{\mathcal{Q}}_{\rm v} &=& 0, \nonumber 
\ee
where we have used $\left\langle \delta \mathbf{B}\right\rangle = 0$,
and defined $(\mathbf{B}_{\rm reg})_{x} = B_{\rm reg} \cos \theta_{\rm
  reg}$ and $(\mathbf{B}_{\rm reg})_{y} = B_{\rm reg} \sin \theta_{\rm
  reg}$.  We also note that for $\bar{\Delta}/N \gg 1$ and $N \gg 1$, the variance of the $\mathcal{Q}_{i}$ at a given frequency are $\Oo(\Pbar^2)$. We present the detailed
evaluation of $\Pphi$ and the $\mathcal{Q}_{i}$ in the cell model in
Appendix \ref{app:Cell}. These results were first presented in Ref. \cite{Burrage08}.

We note that if, as is generally the case, $B_{\rm rand} \sim \Oo(B_{\rm reg})$, the effect of multiple box-like magnetic regions with coherence length 
$L_{\rm coh} \ll L$ is to enhance the expected photon-ALP mixing probability by a factor of $N$.

\subsection{The Power Spectrum Model} \label{sub:PowerSpectrumModel}
A more generalised prescription for describing the magnetic field and
 electron density fluctuations is with a power spectrum model. As before, $m_{\rm eff}^2 \simeq
 -\omega_{\rm pl}^2 \propto n_{\rm e}$.  We assume that, for each $i$
 and for fixed $\mathbf{x}$, the turbulent component of the magnetic
 field $\delta \mathbf{B}_{i}(\mathbf{x})$ is approximately a Gaussian
 random variable. We assume that the fluctuations are isotropic and so
 $$R_{{\rm B}\,ij}(\mathbf{x}; \mathbf{y}) \equiv \left\langle \delta
 \mathbf{B}_{i}(\mathbf{y})\delta
 \mathbf{B}_{j}(\mathbf{x}+\mathbf{y})\right\rangle =
 \frac{1}{3}R_{\rm  B}(\mathbf{x};\mathbf{y}) \delta_{ij}.$$ Finally
 we require (at least approximate) position independence for the
 fluctuations so that $R_{\rm  B}(\mathbf{x};\mathbf{y}) \approx
 R_{\rm B}(x)$. $R_{\rm B}(x)$ is the auto-correlation function for
 the magnetic field fluctuations. The electron number density is divided into a constant part and a fluctuating part, $n_{\rm e} = \bar{n}_{\rm e} + \delta
 n_{\rm e}$, and we assume that $1+\delta n_{\rm e}/\bar{n}_{\rm e}$  is a
 log-normally distributed random variable with mean $1$ and variance
 $\left\langle \delta_{\rm n}^2 \right\rangle$, where $\delta_{\rm
 n}\equiv \delta n_{\rm e}/\bar{n}_{\rm e}$.  We define the electron
 density auto-correlation function by, $$R_{\rm N}(x) = \left\langle \delta n_{\rm e}(\mathbf{y})\delta n_{\rm e}(\mathbf{x}+\mathbf{y})\right\rangle.$$  

The power spectra for the  magnetic and electron density fluctuations, $P_{\rm B}(k)$ and $P_{\rm N}(k)$, are defined:
\be
R_{\rm B}(x) &=& \frac{1}{4\pi} \int \dd^3 k e^{2\pi i \mathbf{k}\cdot\mathbf{x}} P_{\rm B}(k) \nonumber \\
&=& \int k^2 \dd k \,P_{\rm B}(k) \frac{\sin (2\pi kx)}{2\pi k x}, \nonumber \\
R_{\rm N}(x) &=& \frac{1}{4\pi} \int \dd^3 k e^{2\pi i \mathbf{k}\cdot\mathbf{x}} P_{\rm N}(k) \nonumber \\
&=& \int k^2 \dd k \,P_{\rm N}(k) \frac{\sin (2\pi kx)}{2\pi k x}. \nonumber 
\ee 
Following Ref. \cite{Murgia04}, we define correlation length scales, $L_{\rm B}$ and $L_{\rm N}$, for the magnetic and electron density fluctuations respectively:
\be
L_{\rm B/N} = \frac{ \int_{0}^{\infty} k\dd k \,P_{\rm B/N}(k)}{2 \int_{0}^{\infty} k^2\dd k \, P_{\rm B/N}(k)}. \label{lengths}
\ee
We define $\bar{\Delta} = \bar{m}_{\rm eff}^2 L/4\omega \simeq -4\pi
\alpha_{\rm em} \bar{n}_{\rm e} L /4m_{\rm e} \omega$, and $k_{\rm
  crit} = \vert\bar{\Delta} /\pi L\vert$.  In Appendix
\ref{app:Powerspectrum} we evaluate $\Pbar$ at low frequencies,
i.e. large $\vert \bar{\Delta}\vert$, in the power spectrum model. We
assume there is some $k_{-3}^{(\rm B/N)} \ll k_{\rm crit}$ such that
$P_{\rm B/N}(k)$ decreases faster than $k^{-3}$ as $k \rightarrow
\infty$ for $k > k_{-3}^{(\rm B/N)}$. This is equivalent to assuming
that the dominant contribution to $\left\langle \delta
\mathbf{B}^2\right\rangle$ and $\left\langle n_{\rm e}^2
\right\rangle$ comes from spatial scales than are much larger than
$k_{\rm crit}^{-1}$.  We expect that this dominant contribution will
come from scales of the order of the coherence lengths of the magnetic
field and electron density fluctuations and so are assuming that $k_{\rm crit}^{-1} \ll L_{\rm
  B},\,L_{\rm N}$.   This and the other assumptions given above lead to,
\be
\Pbar &\equiv& \left \langle \Pphi \right\rangle
\approx\frac{1}{2}\left(\frac{2B_{\rm eff} \omega}{M_{\rm eff}
  \bar{m}_{\rm eff}^2}\right)^2 I_{\rm N}^3 \label{Ppower}\\ &&-
\frac{1}{4}\left(\frac{2B_{\rm reg} \omega}{M_{\rm eff} \bar{m}_{\rm
    eff}^2}\right)^2 \cos \left(2\bar{\Delta}\right) \nonumber \\ && + \frac{B_{\rm eff}^2 L }{8M_{\rm
  eff}^2 \bar{n}_{e}^2} I_{\rm N}^2 W_{\rm N}(k_{\rm crit}) +
\frac{L}{24M_{\rm eff}^2} I_{\rm N}^3  W_{\rm B}(k_{\rm crit}). \nonumber
\ee
where
\be
I_{\rm N} &=& 1+\left\langle \delta_{\rm n}^2\right\rangle,\nonumber \\
B_{\rm eff}^2 &=& \frac{1}{2}\left\langle (\hatb{z} \times \mathbf{B})^2 \right\rangle \nonumber \\
&=& B_{\rm reg}^2/2 +  \left\langle \delta \mathbf{B}^2\right\rangle/3,\nonumber \\
W_{{\rm B/N}}(k_{\rm crit}) &=& \int_{k_{\rm crit}}^{\infty} k \dd k \, P_{\rm B/N}(k).\nonumber
\ee
It is straightforward to check that $\bar{\mathcal{Q}}_{\rm v} = 0$, and
\be
\bar{\mathcal{Q}}_{\rm q} &\approx& \mathcal{Q}_{0}\cos 2\theta_{\rm reg}, \nonumber \\
\bar{\mathcal{Q}}_{\rm u} &\approx& \mathcal{Q}_{0}\sin 2\theta_{\rm reg}, \nonumber
\ee
where
\be
\mathcal{Q}_{0} &=& \frac{1}{4}\left(\frac{2B_{\rm reg} \omega}{M_{\rm
    eff}\bar{m}_{\rm eff}^2}\right)^2 I_{\rm N}^3 \nonumber \\ &&-
\frac{1}{4}\left(\frac{2B_{\rm reg} \omega}{M_{\rm eff} \bar{m}_{\rm
    eff}^2}\right)^2 \cos \left(2\bar{\Delta}\right) \nonumber \\ && + \frac{B_{\rm reg}^2 L
}{16M_{\rm eff}^2 \bar{n}_{\rm e}^2} I_{\rm N}^2 W_{\rm N}(k_{\rm crit}). \nonumber
\ee
We note that, in contrast to the cell model, here we have an extra term in the expressions for the mean induced Stokes parameters which scales as $L$. This term is associated with electron density fluctuations and is usually dominant when $\bar{\Delta} \gg 1$, as is the case for CMB photons in galaxy clusters.

We have assumed that near $k=k_{\rm crit}$ and for all large $k$, both $P_{\rm B}(k)$ and  $P_{\rm N}(k)$ are decreasing faster than $k^{-3}$. We further assume that near 
$k = k_{\rm crit}$, $P_{\rm B}(k) \propto P_{\rm N}(k) \propto k^{\alpha-2}$ for some $\alpha < -1$. We define normalization constants $C_{\rm K}$ and $C_{\rm N}$:
\be
k^2 P_{\rm B}(k) &=& 2C_{\rm K}\left(\frac{k}{k_{0}}\right)^{\alpha}, \label{Ck}\\
k^2 P_{\rm N}(k) &=& 2(2\pi)^{1/3} C_{\rm N}^2 k^{\alpha}, \label{Cn}
\ee
where $k_{0} = 1\kpc^{-1}$. These forms hold for $k \sim O(k_{\rm crit})$. We have chosen the normalization for $P_{\rm B}$ and $P_{\rm N}$ to be consistent with the conventions 
of Refs. \cite{Han04} and \cite{Armstrong95} respectively, where $\alpha =-5/3$ corresponds to three-dimensional Kolmogorov turbulence. It follows that,
\be
W_{\rm B}(k_{\rm crit}) &\approx & \frac{2C_{\rm K}}{\vert \alpha \vert} \left(\frac{k_{\rm crit}}{k_{0}}\right)^{\alpha}, \nonumber \\
W_{\rm N}(k_{\rm crit}) &\approx & \frac{2(2\pi)^{1/3}}{\vert \alpha \vert}  C_{\rm N}^2 k_{\rm crit}^{\alpha}. \nonumber
\ee
We have
\be
k_{\rm crit}^{-1} \approx 2.4 \times 10^{-2}\pc \left(\frac{\nu}{100\,{\rm GHz}}\right) \left(\frac{10^{-3}\,{\rm cm}^{-3}}{\bar{n}_{\rm e}}\right), \label{kcrit}
\ee
where $\nu = \omega/2\pi$ is the frequency of the electromagnetic
radiation.  For galaxy clusters  $\bar{n}_{\rm e} \sim 10^{-3}\textrm{--} 10^{-2}\,{\rm cm}^{-3}$ and for CMB photons $\nu \sim
30-300\,{\rm GHz}$. The typical critical length scale is then much smaller than a parsec, $k_{\rm crit}^{-1} 
\approx 10^{-3}\textrm{--}0.1\,{\rm pc}$. The form and magnitude of $P_{\rm B}(k)$ and $P_{\rm N}(k)$ in galaxy clusters for such small scales are not known empirically.

We note that the cell model is in fact a sub-class of the power spectrum model with an appropriate choice for $W_{\rm B}$ and $W_{\rm N}$. When referring to the power spectrum model in what follows we will be referring to the specific case of $P_{\rm B}(k) \propto P_{\rm N}(k) \propto k^{\alpha-2}$ at large $k$.

\section{Magnetic Fields in Galaxy Clusters}\label{sec:ClusterMagneticField}
We saw in the previous section that the mixing of photons with light scalars depends crucially on the structure of the magnetic field along the line of sight.  As the CMB photons propagate towards Earth they pass through clusters
of galaxies. The space between individual galaxies in the cluster
is filled with a hot plasma which supports a magnetic field. In this article, we are focused on the mixing of CMB photons and light (chameleonic) scalars in the intra-cluster (IC) medium.  In this section we briefly review the observed properties of IC magnetic fields.
 
Intra-cluster magnetic fields have been observed to have field strengths
as high as $30\,\mu\hbox{G}$ \cite{Taylor93}. The galaxy clusters can
be divided into two types: non-cooling core clusters for which the
magnetic field in the ICM is generally found to be
a few to $10\,\mu\mathrm{G}$ and ordered on scales of $10-20\,\mathrm{kpc}$;
while in cooling core systems the magnetic field strength is generally
higher (around $10-40\,\mu\mathrm{G}$) and ordered on smaller scales
of a few to $10\,\mathrm{kpc}$ \cite{Clarke04}.  In Table~\ref{tab:galaxy-parameters}
we list the typical ranges for some of the galaxy cluster properties.
\begin{table*}
\begin{centering}
\begin{tabular}{|l|l|}
\hline 
Features of a typical galaxy cluster : & \tabularnewline
\hline 
magnetic field strength, $\left\langle \mathbf{B}^2\right\rangle^{1/2}$ & $0.1$ - $30\,\mu\hbox{G}$,\tabularnewline
coherence length of magnetic field, $L_{\rm coh}$ & $1$ - $100\hbox{\, kpc}$,\tabularnewline
number of magnetic domains, $N=L/L_{\rm coh}$  & $100$ - $10^{3}$ \tabularnewline
electron number density, $n_{\rm e}$ & $10^{-4}$ - $10^{-2}\hbox{\, cm}^{-3}$,\tabularnewline
\hline
\end{tabular}
\par\end{centering}

\caption{\label{tab:galaxy-parameters}List of the parameters for typical
galaxy clusters \cite{Kronberg94,Ohno03}.}

\end{table*}

One of the common methods for determining cluster
magnetic fields is the observation of Faraday rotation measures (RMs).
The RM corresponds to a rotation of the polarization angle of an intrinsically
polarized radio source located behind or embedded within the galaxy
cluster. The RM along a line of sight in the $\hatb{z}$ direction of a source at $z=z_{\rm s}$ is given by:
\be
{\rm RM}(z_{\rm s}\hatb{z}) = a_{0}\int_{0}^{z_{\rm s}} n_{\rm e}(x \hatb{z})B_{\parallel}(x \hatb{z}) \dd x, \nonumber
\ee
where $a_{0} = \alpha_{\rm em}^3/\pi^{1/2} m_{\rm e}^2\,$, $B_{\parallel}$
is the magnetic field along the line of sight, and the observer is at
$\mathbf{x}=0$. For a uniform magnetic field over a length $L$ along the line of sight and in a region of constant density, we expect $\langle {\rm RM}\rangle=a_{0} B_{\parallel}n_{\rm e}L$. However, the dispersion of Faraday RMs from extended
radio sources suggests that this is not a realistic model for IC
magnetic fields. Both the dispersion of the measured RMs and simulations suggest that IC magnetic fields typically have a sizeable component that is tangled on scales smaller than the cluster. The simplest model for a tangled magnetic field
is, as we mentioned in the previous section, the cell model where the
cluster is divided up into uniform cells with sides of length $L_{\rm
  coh}$.  The total magnetic field strength is assumed to be the same
in every cell, but the orientation of the magnetic field in each cell
is random. In the cell model the RM is given by a
Gaussian distribution with zero mean, and variance
\be
\left \langle {\rm RM}^2 \right \rangle \sim \frac{a_{0}^2 L_{\rm coh}} {2} \int_{0}^{L} \left \langle \mathbf{B}^2(z)\right\rangle n_{\rm e}^2(z) \dd z,\nonumber
\ee
where $\left \langle \mathbf{B}^2\right\rangle$ is the average value of $\mathbf{B}^2$ along the line of sight.
We noted in the previous section that a more realistic model for the magnetic field is to assume a power spectrum of fluctuations.  In this model one finds \cite{Murgia04},
\be
\left \langle {\rm RM}^2 \right \rangle \sim \frac{a_{0}^2  L_{\rm B} }{2} \int_{0}^{L} \left \langle \mathbf{B}^2(z) \right\rangle n_{e}^2(z) \dd z\,. \nonumber
\ee
To match the same observations with the same total field strength, $L_{\rm B}$ in the power spectrum model should be equal to $L_{\rm coh}$ in the cell model. Whichever model is used, determination of the magnetic field from Faraday rotation measures
is heavily reliant on the estimated coherence length of the magnetic
field. A study by Clarke \emph{et al.} \cite{Clarke01} examining the Faraday
RMs of 16 ``normal'' clusters, i.e. free of strong radio halos
and widespread cooling flows, found the following relationship between
the average field strength and the inferred coherence length for the IC
magnetic field:
\begin{equation}
\langle\left|B\right|\rangle_{\mathrm{icm}}=5\text{--}10\,\left(\frac{L_{\mathrm{dom}}}{10\,\mathrm{kpc}}\right)^{-1/2}h_{75}^{1/2}\,\mu\mathrm{G}\label{eq:B-L relation}
\end{equation}
where $h_{75}$ is defined by $H_{0}=75\,h_{75}\,\mathrm{km\, s^{-1}Mpc^{-1}}$.
This study excludes the cooling core systems which generally
possess slightly higher magnetic field strengths. 

The ambient electron number density also plays a crucial role in determining the strength of photon-scalar mixing. 
The electron density of cluster atmospheres can be determined from
X-ray surface brightness observations, for example using ROSAT, which
are fitted to a $\beta$ profile (see Ref. \cite{Govoni04}): $$n_{e}(r)=n_{0}\left(1+\frac{r^{2}}{r_{c}^{2}}\right)^{-3\beta/2},$$
where $\beta\sim\mathcal{O}(1)$ and positive. The magnetic field
is expected to decline with radius in a similar fashion to the electron
density. Magneto-hydrodynamic (MHD) simulations of
galaxy cluster formation suggest that $ \left \langle \mathbf{B}^2 \right\rangle^{1/2} \propto \left \langle n_{e} \right \rangle^{\eta}$, $\eta=1$
while observations of Abell 119 suggest $\eta=0.9$ \cite{Dolag}.
Two simple theoretical arguments for the value of $\eta$, are (1) that the magnetic field is frozen-in and hence $\eta = 2/3$, or (2) the total magnetic energy, $\propto \left\langle \mathbf{B}^2\right\rangle$, scales as the thermal
energy, $\propto n_{\rm e}$, and hence $\eta = 1/2$ \cite{Govoni04}. Faraday RM observations of clusters \cite{Clarke01} have found that the magnetic field extends as far as the ROSAT-detectable
X-ray emission (coming purely from the electron density distribution). Typically this corresponds to a scale of $\sim 500\,\mathrm{kpc}$.

\subsection{Power Spectra of Cluster Magnetic Fields}\label{sub:PowerSpectra}
As discussed in \S \ref{sub:PowerSpectrumModel}, fluctuations across a range of scales in the magnetic field and electron density in clusters can be described by their respective power spectra, $P_{\rm B}(k)$ and $P_{\rm N}(k)$. On small scales the power spectra are parameterised by a power law form as given by Eqs. (\ref{Ck}) and (\ref{Cn}). In these equations the index $\alpha =-5/3$ corresponds to the special case of three dimensional Kolmogorov turbulence. According to Kolmogorov's 1941 theory of turbulence, the power spectrum of three dimensional turbulence on the smallest scales is universal and has $P(k) \propto k^{-11/3}$. 

The values of the normalization constants, $C_{\rm N}^2$ and $C_{\rm K}$, in our own galaxy have been determined by Armstrong \etal  \cite{Armstrong95} and Minter \& Spangler \cite{Minter96} respectively.   Armstrong \etal  found $\alpha \approx -5/3$ for spatial scales $10^{6}{\rm m} < k^{-1} < 10^{13}{\rm m}$, and $C_{\rm N}^2 \approx 10^{-3}{\rm m}^{-20/3}$.  Minter \& Spangler also found $\alpha \approx -5/3$ for spatial scales $0.01 \,{\rm pc} \lesssim k^{-1} \lesssim 3.6\,{\rm pc}$, and $C_{\rm K} \approx 0.95\times 10^{-12}\,{\rm erg}\,{\rm cm}^{-3}\kpc$.  On larger scales  Han \etal \cite{Han04} found that the magnetic fluctuation power spectrum flattens with  $\alpha \approx -0.37 \pm 0.1$ for $0.5 \kpc \lesssim k^{-1} \lesssim 15\kpc$, and $C_{\rm K} \approx 6.8 \times 10^{-13}\,{\rm erg}\,{\rm cm}^{-3}\kpc$ on these scales.

En{\ss}lin \& Vogt \cite{Ensslin03} and Murgia \etal \cite{Murgia04} showed that information about the power
spectrum of galaxy cluster magnetic fluctuations could also be gained from detailed RM images. The inferred galaxy cluster magnetic power spectra are consistent with a power law form and $-2< \alpha <
0$ in different regions of the clusters.  For instance, in Ref. \cite{Vogt03} Vogt \& En{\ss}lin analysed Faraday rotation maps of three galaxy clusters: Abell 400, Abell 2634 and Hydra A.  They found that for $k \gtrsim 1\kpc^{-1}$, the power spectra of all three clusters were consistent with a power law form and $-2 < \alpha < -1.6$, and hence consistent with a Kolomogorov power spectrum ($\alpha = -5/3$) on small scales.  For $k \lesssim 1\kpc^{-1}$, $k^2 P_{\rm B}$ was found to be roughly flat or slightly increasing with $k$.

Such measurements only probe the power spectrum at spatial scales larger than a few kiloparsecs.  In this work, we find that the photon to scalar field conversion rate is 
sensitive to $P_{\rm B}(k)$ and $P_{\rm N}(k)$ on scales $\sim k_{\rm crit}^{-1}$ (Eq. (\ref{kcrit})). Typically $k_{\rm crit}^{-1} \approx 10^{-3}\textrm{--}0.1\,{\rm pc}$ for the CMB in galaxy clusters.  There are no observations of the form or magnitude of $P_{\rm B}(k)$ and $P_{\rm N}(k)$ on such small scales. We therefore estimate 
$\alpha$ by assuming that on scales $\lesssim k_{\rm crit}^{-1}$, we have three dimensional Kolmogorov turbulence.   The dominant contribution to 
$\left\langle \delta \mathbf{B}^2 \right\rangle$ comes from scales where $P_{\rm B}$ switches from dropping off more slowly than $k^{-3}$ to dropping off faster than $k^{-3}$ as 
$k$ increases. We define this scale to be $k_{\ast}$.   To estimate  $C_{\rm K}$ and $C_{\rm N}$ we assume a simple model for $P_{\rm B/N}(k)$ whereby for $k \gtrsim k_{\ast}$ up
 to and including $k_{\rm crit}$, we have Kolmogorov turbulence, and for $k \lesssim k_{\ast}$, the power spectrum decreases more slowly than $k^{-3}$ with increasing $k$. We also
  assume that $k^2 P_{\rm B/N}(k) \rightarrow {\rm const}$ as $k \rightarrow 0$ so that the coherence lengths $L_{\rm B}$ and $L_{\rm N}$ (Eq. (\ref{lengths})) are well defined.  We therefore assume,
\be
k^2 P_{\rm B}(k) &\approx& 2C_{\rm K} \left(\frac{k}{k_0}\right)^{-5/3}, \quad k > k_{\ast}, \nonumber\\
k^2 P_{\rm B}(k) &\approx& 2C_{\rm K} \left(\frac{k_{\ast}}{k_0}\right)^{-\frac{5}{3}}\left(\frac{k}{k_\ast}\right)^{\beta}, \quad k_{\rm min} < k < k_{\ast}, \nonumber
\ee
where $\beta > -1$, and we can ignore the contribution of $k^2 P_{\rm B}(k)$ for $k < k_{\rm min}$. It follows that,
\be
\left\langle \delta \mathbf{B}^2\right\rangle &=& \frac{1}{4\pi} \int \dd^3 k P_{\rm B}(k) \nonumber \\
&=& \left[2f_{1+\beta}(x)+3\right] C_{\rm K} k_{\ast} \left(\frac{k_{\ast}}{k_0}\right)^{-5/3}, \nonumber \\
L_{\rm B}\left\langle \delta \mathbf{B}^2\right\rangle &=& \frac{1}{2}\int k\,\dd k\, P_{\rm B}(k) \nonumber \\
&=& \left[f_{\beta}(x) + \frac{3}{5}\right] C_{\rm K}\left(\frac{k_{\ast}}{k_0}\right)^{-5/3}, \nonumber
\ee
where $x=k_{\rm min}/k_{\ast}$ and we have defined
$$
f_{\beta}(x) = \frac{1}{\beta}\left[1-x^{\beta}\right].
$$
From which we find,
\be
k_{\ast} &=& L_{\rm B}^{-1}\left(\frac{f_{\beta}(x) + \frac{3}{5}}{2f_{1+\beta}(x)+3}\right), \nonumber 
\ee
and
\be
k_{0}^{5/3}C_{\rm K} &=& \frac{4}{5^{7/3}} h(x)\left\langle \delta \mathbf{B}^2 \right\rangle L_{\rm B}^{-2/3}, \nonumber
\ee
where
$$
h(x) \equiv \frac{5^{7/3}(f_{\beta}(x) + \frac{3}{5})^{2/3}}{4(2f_{1+\beta}(x)+3)^{5/3}}.
$$
On large scales, $100\,{\rm kpc}^{-1} \lesssim k \lesssim 10 \,{\rm kpc}^{-1}$, observations such as those in Ref. \cite{Vogt03} are consistent with $\beta \approx 0$. We approximate $k_{\rm min} \sim O(L_{\rm clust}^{-1})$ where $L_{\rm clust}$ is the length scale of the magnetic region and we have
$k_{\ast} \sim O(L_{\rm B}^{-1})$. Thus  $k_{\ast}/ k_{\rm min} \sim 10 - 200$ which gives, taking $\beta \approx 0$,
\be
h(x) \approx \left(\frac{5}{8}\ln (k_{\ast}/k_{\rm min}) + \frac{3}{8}\right)^{2/3} \sim 1.5 \textrm{--} 2.6. \nonumber
\ee
Hence,
\be
2k_{0}^{5/3}C_{\rm K}  \approx \left(0.27 \textrm{--} 0.45\right)\left\langle \delta \mathbf{B}^2 \right\rangle L_{\rm B}^{-2/3}. \nonumber
\ee
For the galactic magnetic field where $\left\langle \delta \mathbf{B}^2 \right\rangle^{1/2} \approx 3\,\mu{\rm G}$, and $L_{\rm B} \sim 20\textrm{--}50\,{\rm pc}$, this gives
\be
C_{\rm K} \sim (0.7 \textrm{--} 2.2) \times 10^{-12}\,{\rm erg}\,{\rm cm}^{-3}\kpc, \nonumber
\ee
in line with observations. Following a similar procedure for $C_{\rm N}^2$ we estimate,
\be
2(2\pi)^{1/3}C_{\rm N}^2 \bar{n}_{e}^{-2} \approx \left(0.27 \textrm{--} 0.45\right)\frac{\left\langle\delta n_{\rm e}^2\right\rangle}{\bar{n}_{\rm e}^2} L_{\rm B}^{-2/3}, \nonumber
\ee
where $\left\langle\delta n_{\rm e}^2\right\rangle/\bar{n}_{\rm e}^2 =I_{\rm N}-1$, and we approximate $L_{\rm N}$ by $L_{\rm B}$. The above expressions provide order of magnitude estimates for $C_{\rm K}$ and $C_{\rm N}^2$.

\section{Chameleonic corrections to the Sunyaev--Zel'dovich Effect}\label{sec:SZ}
One of the well-documented foreground effects acting on the CMB as
it passes through galaxy clusters is the Sunyaev--Zel'dovich (SZ) effect.
It arises from inverse-Compton scattering of CMB photons with free
electrons in a hot plasma. The scattering process redistributes the
energy of the photons causing a distortion in the frequency spectrum
of the intensity. This is observed as a change in the apparent brightness
of the CMB radiation in the direction of galaxy clusters.   If a
chameleon field or other light scalar couples to photons, the flux of
CMB photons passing through galaxy clusters would additionally be
altered by the mixing of this scalar field with CMB photons. As we
shall see, for this new effect to be detectable the required
effective scalar-photon coupling, $g_{\rm eff}$, and
effective scalar mass, $m_{\phi}$, are ruled out for standard axion-like-particles.
A standard ALP with the required properties would violate constraints
on ALP production in the Sun, He burning stars and SN1987A.  However,
all of these constraints feature ALP production in high density
regions. If the ALP is a chameleon-like scalar field, then constraints
from high density regions only very weakly constrain the properties of
the scalar field in low density regions such as the ICM. A
chameleon-like scalar field includes the chameleon and OP models
discussed earlier or any other non-standard ALP for which one or both
of $g_{\rm eff}$ and $1/m_{\phi}$ decrease strongly enough as the
ambient density increases. We refer to the correction to the
Sunyaev--Zel'dovich effect due to mixing with some (chameleon-like) ALP as the chameleonic SZ (CSZ) effect.

In this section, we briefly review the properties of the CMB and the standard thermal Sunyaev--Zel'dovich effect before examining the chameleonic  SZ effect.

\subsection{The CMB}\label{sub:The-CMB}
The Cosmic Microwave Background (CMB) was emitted in the early Universe
from the surface of last scattering at the time of recombination.
This radiation is approximately uniform across the sky and has the
intensity spectrum of a black-body radiating at a temperature of $T_{0}=2.725K$:
\be
I_{0}=\frac{k_{\rm B}T_{0}}{2\pi^{2}}\frac{\omega^{2}}{c^{2}}\frac{x}{e^{x}-1}\,, \nonumber
\ee
where $x\equiv \omega/k_{\rm B}T_{0}$, and $k_{\rm B}$ is the Boltzmann constant. In addition to
this monopole there are small fluctuations in the primary CMB arising
from primordial density fluctuations at the time of emission. Small fluctuations in intensity are related to the changes in the
characteristic black-body temperature by
\be
\frac{\delta I}{I_{0}}&=& \frac{x}{1-e^{-x}}\frac{\delta T}{T_{0}}.\nonumber
\ee
The root mean squared average of these primary fluctuations is $\Oo(10^{2})\mu {\rm K}$.  As the CMB radiation
propagates towards Earth there are foreground effects modifying the
intensity which give rise to secondary fluctuations in the CMB. We
expect the chameleonic SZ effect in galaxy clusters to be one such foreground.
When considering these foreground effects, one generally treats the
primary CMB fluctuations as Gaussian noise on top of the CMB monopole
signal.

The primordial CMB radiation is also weakly polarized. The intensity
and polarization of radiation in general is described by its Stokes
vector $(I,\, Q,\, U,\, V)$. The fractional linear polarization
of the CMB is of the order $\langle Q^{2}\rangle^{1/2}=\langle U^{2}\rangle^{1/2}\sim10^{-6}$,
while the circular polarization component, $V$, is predicted to be zero
from parity considerations in the early Universe.

\subsection{The Thermal SZ Effect}\label{sub:ThermalSZ}
In the standard SZ effect the predicted fractional change to
the temperature of CMB photons passing through a galaxy cluster is \cite{Birkinshaw99}
\be
\frac{\Delta T_{\mathrm{SZ}}}{T_{0}}=\frac{k_{\rm B}T_{\rm e}}{m_{\rm e}}\tau_{0}\left(x\coth\left(\frac{x}{2}\right)-4\right)\,,\label{eq:SZ effect}
\ee
where $x=\omega/ k_{\rm B}T_{0}$, $T_{\rm e}$ is the temperature of the electrons in the plasma,
and $\tau_{0}\equiv\int\sigma_{\rm T}n_{\rm e}(l)\mathrm{d}l$ is the optical
depth given by the integral along the line of sight of the number
density of electrons multiplied by the Thompson cross-section, $\sigma_{\rm T} = 6.65 \times 10^{-29}{\rm m}^2$.
At low frequencies the SZ effect is seen as a decrement in the expected
temperature from the CMB, while at high frequencies it is seen as a
boost in the temperature. There are relativistic corrections to the
above formula, but for the specific case of the Coma cluster that
we consider in this paper the peculiar velocity is small and higher
order corrections to the thermal SZ effect are negligible \cite{Itoh}.

\subsection{The Chameleonic SZ Effect}\label{sec:CSZ}
We found in \S \ref{sec:Optics} that on passing through a magnetic
field of length $L$, the photon-scalar mixing alters the intensity of light from $I_{0}$ to $I_{\rm f}$, where $I_{\rm f}$ is given by
\be
I_{\rm f} &=& I_{0}\left(1-\Pphi(L)\right) \nonumber \\ &&+ Q_{0} \mathcal{Q}_{\rm q}(L) + U_{0} \mathcal{Q}_{\rm u}(L) + V_{0} \mathcal{Q}_{\rm v}(0). \nonumber
\ee 
The $Q_{0}$, $U_{0}$ and  $V_{0}$ are the initial values of the
Stokes parameters.  The CMB radiation is only very weakly polarized
and observations confirm $\left\langle
Q_{0}^2\right\rangle^{1/2}/I_{0}$, $\left\langle
U_{0}^2\right\rangle^{1/2}/I_{0} \sim \Oo(10^{-6})$ and $\left\langle
V_{0}^2 \right\rangle^{1/2}/I_{0} \ll 10^{-6}$.  The root mean squared
values of $\Pphi$ and the $\mathcal{Q}_{i}$ are, however, all of similar magnitude, $\mathcal{O}(\Pbar)$. In addition to this
the measurement issues arising from a finite frequency bin, as discussed in Appendix \ref{app:Measurement}, mean that the observed variance
will be significantly reduced. Hence, for CMB photons, we have that
$$
I_{\rm f} \approx I_{0}\left(1-\Pbar(L)\right),
$$
and so the change in intensity due to the chameleonic SZ effect,  $\Delta I_{\rm CSZ}$, is given by 
\be
\frac{\Delta I_{\rm CSZ}}{I_{0}} = -\Pbar(L). \nonumber
\ee
Transforming this to a change in temperature we have,
\be
\frac{\Delta T_{\rm CSZ}}{T_{0}} = \frac{1-e^{-x}}{x} \frac{\Delta I_{\rm CSZ}}{I_{0}} = \frac{e^{-x}-1}{x} \Pbar(L). \nonumber
\ee
where as before $x =\omega / kT_{0}$. We evaluated the expected
value of $\Pphi(L)$ for different models of the IC magnetic field in
\S \ref{sec:Optics} above, and we note that it depends both on
$n_{\rm e}$, $\mathbf{B}$ and frequency $\omega$.  We assume that inside
the cluster $\left\langle \mathbf{B}^2 \right\rangle^{1/2} \propto
\left\langle n_{\rm e}\right\rangle^{\eta}$. This implies for the cell model we have,
\be
\frac{\Delta I_{\rm CSZ}}{I_{0}} \propto n_{\rm e}^{-2(1-\eta)}\omega^{2}, \label{CellDep}
\ee
whereas in the power spectrum model, where
$P_{\rm B}(k) \propto P_{\rm N}(k) \propto k^{\alpha -2}$ for $k \geq
k_{\rm crit}$ (Eq. (\ref{kcrit})) and $\alpha < -1$, we
have to leading order, 
\be
 \frac{\Delta I_{\rm CSZ}}{I_{0}}\propto n_{\rm e}^{2\eta + \alpha}\omega^{-\alpha}, \label{PowerDep}
\ee
for the case $\alpha > -2$. For $\alpha < -2$, we find the above behaviour for higher frequencies, whereas at lower frequencies we find the same leading order 
 scaling as we had in the cell model. A Kolmogorov spectrum for the magnetic turbulence has $\alpha = -5/3$, if we take $\eta = 0.9$ as suggested by observations, 
 then the CSZ effect scales as $n_{\rm e}^{2/15} \omega^{5/3}$. 

For comparison, the thermal SZ effect is proportional to $n_{\rm e}$
and has a non-power law frequency dependence. We  note that for
$\alpha < -1$ and $\eta \leq 1$, $\Delta I_{\rm CSZ}/I_{0}$ always has
a shallower $n_{\rm e}$ dependence than $\Delta I_{\rm SZ}/I_{0}$.  In
principle, the chameleonic SZ effect can be distinguished from the
standard contributions to the SZ effect using both its electron density and frequency dependences. It should be noted, though, that in both Eqs. (\ref{CellDep}) and (\ref{PowerDep}) we have assumed that the magnetic and electron density correlation lengths do not vary as $n_{\rm e}$ varies.  Simulations suggest that the correlation length increases with radius, although the precise form of this variation is uncertain.  If we assume that $L_{\rm coh} = L_{\rm B} \propto L_{\rm N}$, and estimate the high-$k$ power spectrum according to the prescription given in \S \ref{sub:PowerSpectra}, we see that 
$$
 \frac{\Delta I_{\rm CSZ}}{I_{0}} \propto n_{\rm e}^{2\eta + \alpha_{0}}\omega^{-\alpha_{0}} L_{\rm coh}^{1+\alpha_{0}}, 
$$
where $\alpha_{0} = \alpha$ if $-2 \leq \alpha < -1$, and $\alpha_{0} = -2$ in the cell model. The $L_{\rm coh}$ scaling for $\alpha <-2$ is more complicated, at higher
frequencies it is given by the above formula with $\alpha_0 = \alpha$, whereas at lower frequencies, $\Delta I_{\rm CSZ}$ is independent of $L_{\rm coh}$ at leading order. In what
follows we assume that $-2 \leq \alpha < -1$ for the power spectrum model. We discuss how the spatial variation of $\Delta I_{\rm CSZ}$ can be used to constrain photon-scalar mixing further in \S \ref{sec:SZprofiles} below.

\section{Application}\label{sec:Application}
In this section, we apply the results derived above to constrain the probability of photon-scalar mixing at CMB frequencies in the intra-cluster medium.   
We found that in the power spectrum model, $\Pbar \propto \omega^{-\alpha}$, for $-2 \leq \alpha < -1$ where $\alpha$ is the slope of the power spectra for the magnetic and 
electron density fluctuations in the region $k \geq k_{\rm crit}$ (Eq. (\ref{kcrit})). The cell model for the magnetic field effectively has $\alpha = -2$.  We write 
$$
\Pbar(L,\omega) = \Pbar(L,\omega_{0}) \left(\frac{\omega}{\omega_{0}}\right)^{-\alpha},
$$
for some fixed frequency $\omega_{0}$ of the order of the CMB
frequency. For given objects we are able to constrain $\Pbar(L,\omega_{0})$, the average probability of photon to scalar conversion at $\omega_{0}$.  
We pick $\omega_{0}$ so that for the range, $-2 \leq \alpha < -1$, the resulting constraint on $\Pbar(L,\omega)$ is almost independent of $\alpha$.

\subsection{Magnitude of Chameleonic SZ Effect}\label{sub:Mag}
Using our evaluation of $\Pbar(L,\omega)$ in Eqs. (\ref{Pphibox}) and
  (\ref{Ppower}), a constraint on $\Pbar(L,\omega_0)$ can be be
  transformed into a constraint on the effective photon-scalar
  coupling $g_{\rm eff} =1/M_{\rm eff}$.  $\Pbar(L,\omega_0)$ depends
  on quantities which are not known a priori, such as $\left\langle
  \mathbf{B}^2\right\rangle$, $\left\langle n_{\rm e}\right\rangle$,
  $\left\langle n_{\rm e}^2 \right\rangle$, $L_{\rm B}$, $L_{\rm N}$,
  $C_{\rm B}$ and  $C_{\rm N}$. The accuracy to which these parameters can be measured or estimated will therefore limit the constraints on $g_{\rm eff}$. The derived constraint 
  on $g_{\rm eff}$ will also depend strongly on $\alpha$.
  We explicitly consider the relation between $g_{\rm eff}$ and $\Pbar$ for two models: the cell model where effectively $\alpha = -2$ and the Kolmogorov turbulence model 
  where $\alpha = -5/3$. The constraint on $\Pbar(L,\omega_0)$ is, however, reasonably model independent, with only a slight dependence on $\alpha$. 

In the cell model, Eq. (\ref{Pphibox}) gives,
\be
\mathcal{P}^{(\rm cell)}(L,\nu)  &=& \left(3.2 \times 10^{-10}\right)g_{10}^2 B_{10\mu {\rm G}}^2 n_{0.01}^{-2}\nonumber\\ && L_{200} (L^{\rm coh}_{1\kpc})^{-1} \nu_{214}^2\,,\nonumber
\ee
where $B_{10\mu{\rm G}} = \left\langle \delta \mathbf{B}^2\right\rangle^{1/2}/10\mu{\rm G}$, $n_{0.01} = n_{\rm e}/10^{-2}{\rm cm}^{-3}$, $\nu_{214}=\nu/214\,{\rm GHz}$, $L_{200}=L/200\kpc$, $L^{\rm coh}_{1\kpc}=L_{\rm coh}/1\kpc$ and $g_{10} = g_{\rm eff}/(10^{-10}\GeV^{-1})$. We have assumed $\bar{\Delta}/N\gg 1$ and that measurements are over a finite frequency bin so that $\sin^2(\bar{\Delta}/N)\approx 1/2$.  

A potentially more realistic model for the magnetic field structure is to assume that the magnetic and electron density power spectra follow a Kolmogorov power law for $k$ ranging from $L_{\rm coh}^{-1}$ to at least $k_{\rm crit}$, where $k_{\rm crit}$ is given by Eq. (\ref{kcrit}). For $k \lesssim L_{\rm coh}^{-1}$ we assume that $k^{2}P(k)$ is either approximately flat or increasing.  The dominant 
contribution to the magnetic field fluctuations then comes from spatial scales of about $L_{\rm coh}$.  In \S  \ref{sub:PowerSpectra}, we considered a parameterisation of 
$P_{\rm B}(k)$ with these properties, and found
\be
2C_{\rm K}k_{0}^{5/3} = (0.27\text{--}0.45)\left\langle \delta \mathbf{B}^2\right\rangle L_{\rm B}^{-2/3}. \nonumber
\ee
We also assume that the $\delta n_{\rm e}/\bar{n}_{\rm e}$  power spectrum is proportional to that of $\delta \mathbf{B}$.  Under these assumptions, Eq. (\ref{Ppower}) gives,
\be
&\mathcal{P}^{(\rm  Kolmo)}(L,\nu) = \left(2.4\text{--}3.8 \times 10^{-7}\right)g_{10}^2B_{10\mu {\rm G}}^2 n_{0.01}^{-5/3} & \nonumber \\ &\nu_{214}^{5/3}L_{200}\left(L^{\rm coh}_{1 \kpc}\right)^{-2/3} \left(\frac{I_{N}^{2}(2I_{N}-1)}{12}\right). &\nonumber
\ee
A reasonable value for $I_{\rm N}$ based on electron density fluctuations
in our galaxy is $I_{\rm N} \sim 1\text{--}2$, and hence $I_{\rm N}^{2}(2I_{\rm N}-1) \sim 1\text{--}12$. 

The relationship between magnetic field strength and coherence length
found from a study of 16 non-cooling core clusters (Eq. (\ref{eq:B-L relation})) allows us to simplify the above prediction for the case of non-cooling core clusters. For the range  $1\,{\rm kpc}< L_{\rm coh} < 100\,{\rm kpc}$, we find that in the cell model,
\be
8.0 \times 10^{-14} \lesssim \frac{\mathcal{P}^{(\rm cell)}}{g_{10}^2L_{200}n_{0.01}^{-2}\nu_{214}^2} \lesssim  3.2 \times 10^{-9}. \nonumber
\ee
Similarly in the Kolmogorov power spectrum model, including the range for $I_{\rm N}\sim 1\text{--}2$, we find
\be
2.2 \times 10^{-11} \lesssim \frac{\mathcal{P}^{(\rm Kolmo)}}{g_{10}^2L_{200}n_{0.01}^{-5/3}\nu_{214}^{5/3}} \lesssim  3.8 \times 10^{-6}. \nonumber
\ee

Cooling core clusters generally have slightly higher magnetic fields and so we might expect them to exhibit a greater chameleon effect. Hydra A is a good example of a galaxy cluster with strong cooling flows \cite{Taylor93}. Faraday RMs from an embedded radio
source in the cluster have been analysed to suggest a smooth component
to the magnetic field of strength $6\,\mu\mathrm{G}$ ordered on $100\,\mathrm{kpc}$
scales and a tangled component of strength $30\,\mu\mathrm{G}$ ordered
on scales of $4\,\mathrm{kpc}$.  It is this latter component which makes the dominant contribution to $\Pbar$. X-ray data of the cluster indicate a core radius of $R_{\rm core} = 130\,\mathrm{kpc}$ and average electron density of $10^{-2}\,\mathrm{cm^{-3}}$.  We approximate the path length, $L$, as being equal to twice the core radius  $L_{\rm core} = 2R_{\rm core} = 260\,{\rm kpc}$. With these values we find (taking $I_{\rm N} = 1\text{--}2)$,
\be
\mathcal{P}_{\rm Hydra}^{\rm (cell)} &=& \left(0.93 \times 10^{-9}\right)g_{10}^2  \nu_{214}^2, \nonumber \\
\mathcal{P}_{\rm Hydra}^{\rm (Kolmo)} &=& \left(0.90 \text{--} 17.9 \times 10^{-7}\right)g_{10}^2  \nu_{214}^{5/3}. \nonumber
\ee

If, on the other hand, we consider a specific example of a non-cooling core cluster such as the nearby Coma cluster of galaxies, we find little change to the chameleon-photon conversion rate. The IC magnetic field at the centre of the
Coma cluster was determined from Faraday RMs of an embedded
source by Feretti \emph{et al.} \cite{Feretti95}. They found a uniform component of strength
$0.2\pm0.1\,\mu\mathrm{G}$ coherent on scales of $200\,\mathrm{kpc}$
and a tangled component of strength $8.5\pm1.5\,\mu\mathrm{G}$ coherent
on much shorter scales of $1\,\mathrm{kpc}$ extending across
the core region of the cluster ($L_{\rm core}\approx 198\,\mathrm{kpc}$). The electron density at the centre of the cluster is $4 \times 10^{-3}{\rm cm}^{-3}$. We approximate $L$ by $L_{\rm core}$, and find
\be
\mathcal{P}_{\rm Coma}^{\rm (cell)} &=& \left(1.4 \times 10^{-9}\right)g_{10}^2 \nu_{214}^{2}, \label{ComaCell} \\
\mathcal{P}_{\rm Coma}^{\rm (Kolmo)} &=& \left(0.64\text{--}12.7 \times 10^{-7}\right)g_{10}^2 \nu_{214}^{5/3}. \label{ComaKolmo}
\ee
We note that although the Coma cluster core has a weaker magnetic field compared to Hydra A, it also has a lower electron
density which enhances the chameleon effect. Hence the two factors approximately balance out to  give a similar magnitude of the CSZ effect for both Hydra A and Coma.

\subsection{The Coma Cluster} \label{sub:ComaCluster}
\begin{figure}[htb!]
\begin{centering}
\includegraphics[width=7.5cm]{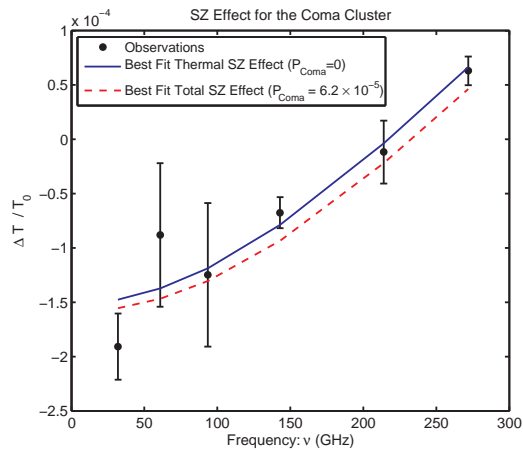}
\end{centering}

\caption{\label{fig:SZ-measurements}SZ measurements of the Coma cluster from
OVRO, WMAP and MITO. The solid line is the predicted thermal SZ effect assuming
a maximum likelihood optical depth, $\tau_{0}=4.7\times 10^{-3}$. The dashed line is the upper bound to the maximum likelihood fit for the combined thermal and chameleonic SZ
effects.}

\end{figure}

The nearby Coma cluster of galaxies is unique in having detailed measurements
of both its magnetic field structure and the SZ effect. 
Using SZ measurements along lines of sight through the core of the Coma cluster, we are able to constrain $\Pphi$ for the Coma cluster. The SZ effect towards the
centre of the cluster has been measured
over a range of frequencies by various probes: OVRO, WMAP and MITO.
These results have been assimilated and analysed in \cite{Batistelli03}.
The frequency measurements of the change to the CMB background temperature
are summarized in Table \ref{tab:ComaSZ} and plotted in Fig \ref{fig:SZ-measurements}. %
\begin{table}
\begin{centering}
\begin{tabular}{|c|c|c|c|}
\hline 
Experiment & $\nu/\mathrm{GHz}$  & $\delta\nu/\mathrm{GHz}$ & $\Delta T_{\mathrm{SZ}}/\mu K$\tabularnewline
\hline
\hline 
OVRO & 32.0 & 6.5 & $-520\pm83$\tabularnewline
\hline 
WMAP & 60.8 & 13.0 & $-240\pm180$\tabularnewline
\hline 
WMAP & 93.5 & 19 & $-340\pm180$\tabularnewline
\hline 
MITO & 143 & 30 & $-184\pm39$\tabularnewline
\hline 
MITO & 214 & 30 & $-32\pm79$\tabularnewline
\hline 
MITO & 272 & 32 & $172\pm36$\tabularnewline
\hline
\end{tabular}
\end{centering}

\caption{\label{tab:ComaSZ}SZ measurements of the Coma cluster.}
\end{table}

\begin{figure}[htb!]
\begin{centering}
\includegraphics[width=7.5cm]{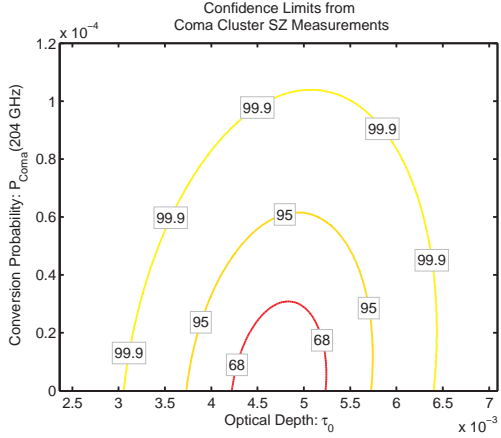}
\end{centering}

\caption{\label{Fig:CLs}Confidence limits from Coma cluster SZ measurements on the optical depth, $\tau_{0}$, and the photon to scalar conversion probability, $\mathcal{P}_{\rm  Coma}$, at $204\,{\rm GHz}$.  }

\end{figure}

The predicted thermal SZ effect was given in Eq. (\ref{eq:SZ effect}) and depends on the optical depth and the
electron temperature in the plasma. For the Coma cluster atmosphere,
$kT_{\rm e}\simeq 8.2\,\mathrm{keV}$. The optical depth $\tau_0 $ is generally treated
as an unknown parameter and inferred from the intensity measurements
using a maximum likelihood analysis. Including the chameleonic contribution which is assumed to scale as $\omega^{-\alpha}$, the predicted form for $\Delta T/T_{0}$ in the Coma cluster is,
\be
\frac{\Delta T^{\rm pred}}{T_{0}} &=& 1.6 \times 10^{-2}\left(x\coth\left(\frac{x}{2}\right)-4\right)\tau_{0} \label{dTpred}\\ &&+ \left(e^{-x}-1\right) x^{-\alpha_{0}-1}
\left(\frac{k_{\rm B}T_{0}}{\omega_{0}}\right)^{-\alpha_{0}} \mathcal{P}_{\rm Coma}(\omega_0), \nonumber
\ee
where $x=\omega/k_{\rm B}T_{0}$; $\alpha_0=\alpha$ if $-2\lesssim \alpha < -1$ in the power spectrum model and $\alpha_0=-2$ in the cell model; and 
$\mathcal{P}_{\rm Coma}(\omega_0) = \Pphi(L,\omega_0)$ for lines

of sight through the centre of the Coma cluster.  We maximize the
likelihood, $L$, over the parameter space of $\tau_{0}$ and $\mathcal{P}_{\rm Coma}(\omega_0)$ where
\be
-2\log L = \sum_{i} \frac{(\Delta T_{i}^{\rm obs}-\Delta T^{\rm pred}_{i}(\tau_0,\mathcal{P}_{\rm Coma}))^2}{\sigma_{i}^2}. \nonumber
\ee
Here $\Delta T_{i}^{\rm obs}$ are the observed values of $\Delta T$ at
a frequency $\nu_{i} = \omega_{i}/2\pi$, and $\sigma_{i}$  are their
standard errors.  $\Delta T^{\rm pred}_{i}$ are the values of $\Delta
T$ as given by Eq. (\ref{dTpred}). We define $\hat{\tau}_{0}$ and
$\hat{\mathcal{P}}_{\rm Coma}$ to be the best fit values of $\tau_{0}$
and $\mathcal{P}_{\rm Coma}$.  Confidence limits are then estimated by
assuming that
\be
\chi^2(\tau_0, \mathcal{P}_{\rm Coma}) = -2\log \left(\frac{L(\tau_0,\mathcal{P}_{\rm Coma})}{L(\hat{\tau}_0,\hat{\mathcal{P}}_{\rm Coma})}\right), \nonumber
\ee
follows a $\chi^2_{1}$ distribution. We find that when $\omega_{0} = 0.844\,{\rm meV}$ ($\nu = 204\,{\rm GHz}$), the constraints on $\mathcal{P}_{\rm Coma}$ are almost independent of $\alpha_{0}$. For all
$-2\leq\alpha_{0}<-1$, the best fit value and estimated standard error of $\tau_{0}$ is virtually independent of $\alpha_{0}$. Fig. \ref{Fig:CLs} shows the $68\%$, $95\%$ and
$99.9\%$ confidence limits on $\tau_{0}$ and $\mathcal{P}_{\rm Coma}(\omega_{0} = 0.844\,{\rm meV})$. Maximizing $\chi^2$ with respect to
${\mathcal{P}}_{\rm Coma}$, gives  $\tau_{0} = (4.7 \pm 1.0) \times 10^{-3}$. Maximizing with respect to $\tau_{0}$ gives the following $95\%$ confidence limit on ${\mathcal{P}}_{\rm Coma}(204\,{\rm GHz})$:
\be
{\mathcal{P}}_{\rm Coma}(204\,{\rm GHz}) < 6.2 \times 10^{-5}\quad (95\%). \nonumber
\ee
The dashed line in Fig. \ref{fig:SZ-measurements} shows the best fit total SZ effect for the $95\%$ confidence upper-bound when $\mathcal{P}_{\rm Coma}(204\,{\rm GHz}) = 6.2 \times 10^{-5}$, with $\alpha_{0}=-2$. It is clear that the preference towards smaller values of $\mathcal{P}_{\rm Coma}$ is due mostly to the measurement points at 
 $143\,{\rm GHz}$ and $272\,{\rm GHz}$ which have the smallest error bars. A very similar picture is seen for other values of $\alpha_{0}$.
From Eqs. (\ref{ComaCell}) and (\ref{ComaKolmo}), we have
\be
&\mathcal{P}_{\rm Coma}^{(\rm cell)}(204\,{\rm GHz})  = \left(1.3 \times 10^{-9}\right)g_{10}^2 \left(\frac{B_{\rm rand}}{8.5\,\mu{\rm  G}}\right)^2& \nonumber \\ & \left(\frac{4 \times 10^{-3}\,{\rm cm}^{-3}}{n_{\rm e}}\right)^2 \left(\frac{L}{198 L_{\rm coh}}\right),&\nonumber \\
&\mathcal{P}_{\rm Coma}^{(\rm  Kolmo)}(204\,{\rm GHz}) = \left(0.59\text{--}11.8 \times 10^{-7}\right)g_{10}^2\left(\frac{B_{\rm rand}}{8.5\,\mu{\rm  G}}\right)^2 & \nonumber\\
&\left(\frac{4 \times 10^{-3}\,{\rm cm}^{-3}}{n_{\rm e}}\right)^{-5/3} \left(\frac{L}{198 \kpc}\right)\left(\frac{L_{\rm coh}}{1 \kpc}\right)^{-2/3}.& \nonumber
\ee
Fixing $B_{\rm rand} = 8.5\,\mu{\rm G}$, $L=198\,{\rm kpc}$, $L_{\rm coh} = 1\,{\rm kpc}$ and $n_{\rm e} = 4 \times 10^{-3}\,{\rm cm}^{-3}$ as suggested by observations, we find that the 95\% confidence upper bound on $\mathcal{P}_{\rm Coma}$ is equivalent to $$g_{\rm eff}^{\rm (cell)} < 2.2\times 10^{-8}\,{\rm GeV}^{-1}$$
in the cell model for the magnetic field, or
$$
g_{\rm eff}^{\rm (Kolmo)} < (7.2 \text{--} 32.5) \times 10^{-10}\,{\rm GeV}^{-1},
$$
if turbulent Kolmogorov power spectrum is assumed.  We note that the constraint $g_{\rm eff} < 7.2 \times 10^{-10}\,{\rm GeV}^{-1}$, which corresponds to 
$I_{\rm N} \equiv 1+ \left\langle \delta n_{e}^2/\bar{n}_{e}^2\right\rangle\approx 2$ and the power spectrum being assumed to be of Kolmogorov form,  is equivalent to 
$M_{\rm eff} > 1.1 \times 10^{9}\GeV$. This was the constraint on $M_{\rm eff}$ found by limiting the chameleonic production of starlight polarization in Ref. \cite{Burrage08}.   It is clear that whilst the constraint on $\Pphi$ at $204\,{\rm GHz}$ is reasonably independent of the magnetic field model, the equivalent constraint on $g_{\rm eff}$ is not and depends strongly on the form and magnitude of the fluctuations in the magnetic field and the electron number density.

\subsection{SZ Profiles}\label{sec:SZprofiles}

The above analysis for the Coma cluster involves measurements of the
average intensity decrement towards the centre of the cluster at multiple
frequencies. Higher frequency measurements are more successful in
constraining the conversion probability than low frequency ones, since $\Delta I_{\rm CSZ}$ grows with frequency.
Potentially better constraints could be found by also considering the
radial profile of the intensity decrement. In \S \ref{sec:ClusterMagneticField}, we discussed
the scaling of the magnetic field and electron density with radius:
\be
\bar{n}_{e} & = & n_{0}\left(1+\frac{r^{2}}{r_{\rm c}^{2}}\right)^{-3\beta/2}, \nonumber \\
\left\langle\mathbf{B}^2 \right\rangle^{1/2} & \propto & \bar{n}_{e}^{\eta}, \nonumber
\ee
where $r_{\rm c}$ is the core radius and we have assumed a $\beta$-model for the gas distribution.   The standard thermal SZ effect is proportional to the integral of $n_{\rm e}$ 
along the line of sight, while the chameleonic SZ effect depends on $n_{\rm e}$, $\left\langle\mathbf{B}^2 \right\rangle^{1/2}$ and $L_{\rm coh}$.  As one moves away from the cluster core, simulations show that the coherence length increases, however the radial scaling $L_{\rm coh}$ is uncertain. To calculate the radial dependence of the chameleonic SZ effect, we must specify how $L_{\rm coh}$ depends on $r$.  We make the simple ansatz that, at least inside the virial radius, $L_{\rm coh} \propto n_{\rm e}^{-\gamma/3\beta}$ for some $\gamma$, so that
\be
L_{\rm coh} &=& L_{{\rm coh}0} \left(1+\frac{r^{2}}{r_{\rm c}^{2}}\right)^{\gamma/2}, \nonumber
\ee
where $L_{{\rm coh}0}$ is the coherence length in the centre of the cluster.  Based solely on dimensional grounds, estimated values for $\gamma$ are: $\gamma = \beta$ so that 
$L_{\rm coh} \propto n_{\rm e}^{-1/3}$, or $\gamma = 1$ so that $L_{\rm coh} \propto r$ for large $r$.  The simulations of Ref. \cite{Ohno03}, suggest that for large $r$, $L_{\rm coh}$ roughly scales as $r$ inside the virial radius, $r_{\rm vir}$.  
\begin{figure*}[htb!]
\begin{centering}
\includegraphics[width = 5cm]{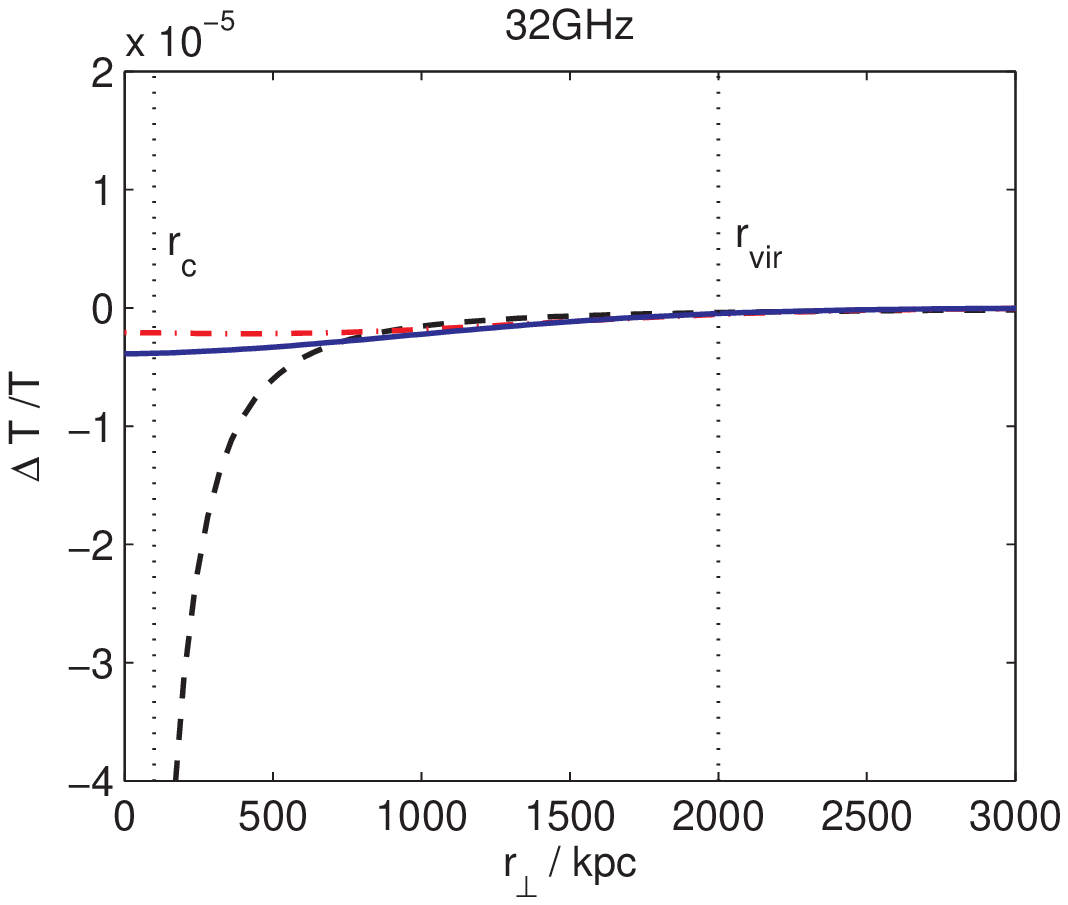}\includegraphics[width = 5cm]{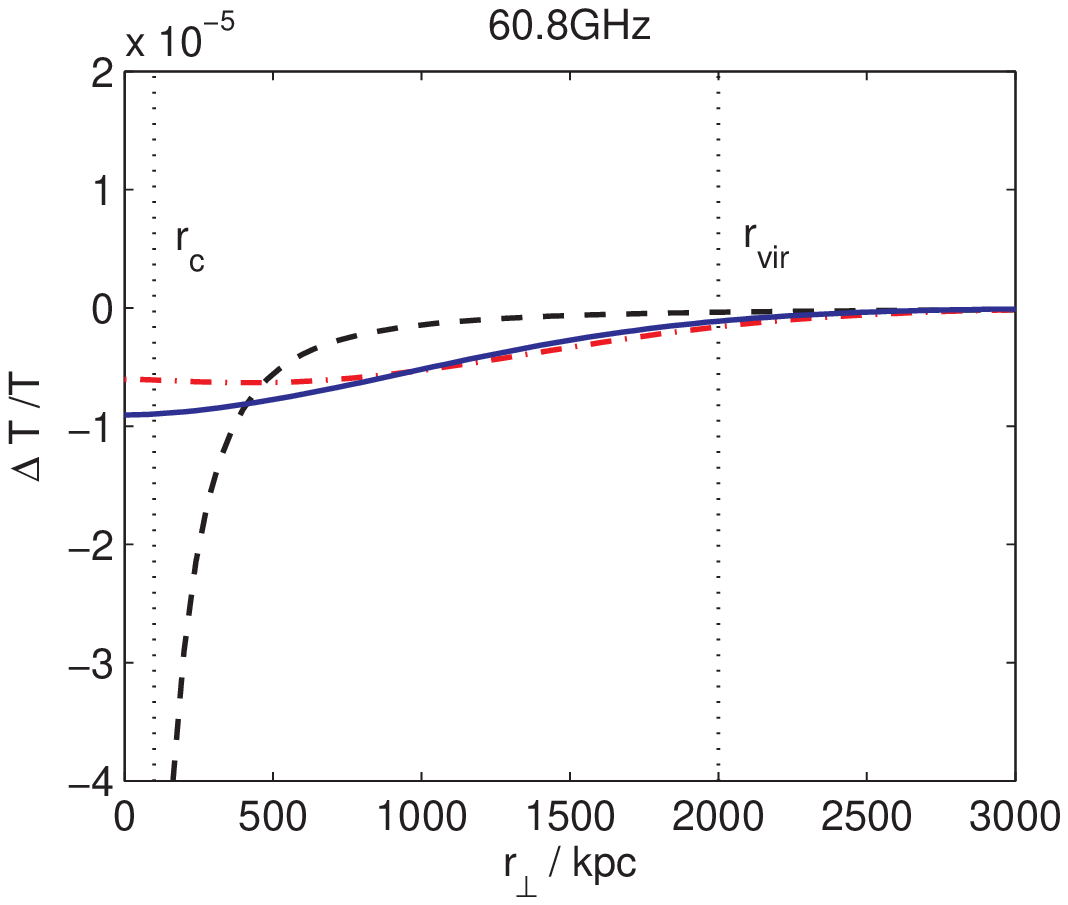}\includegraphics[width = 5cm]{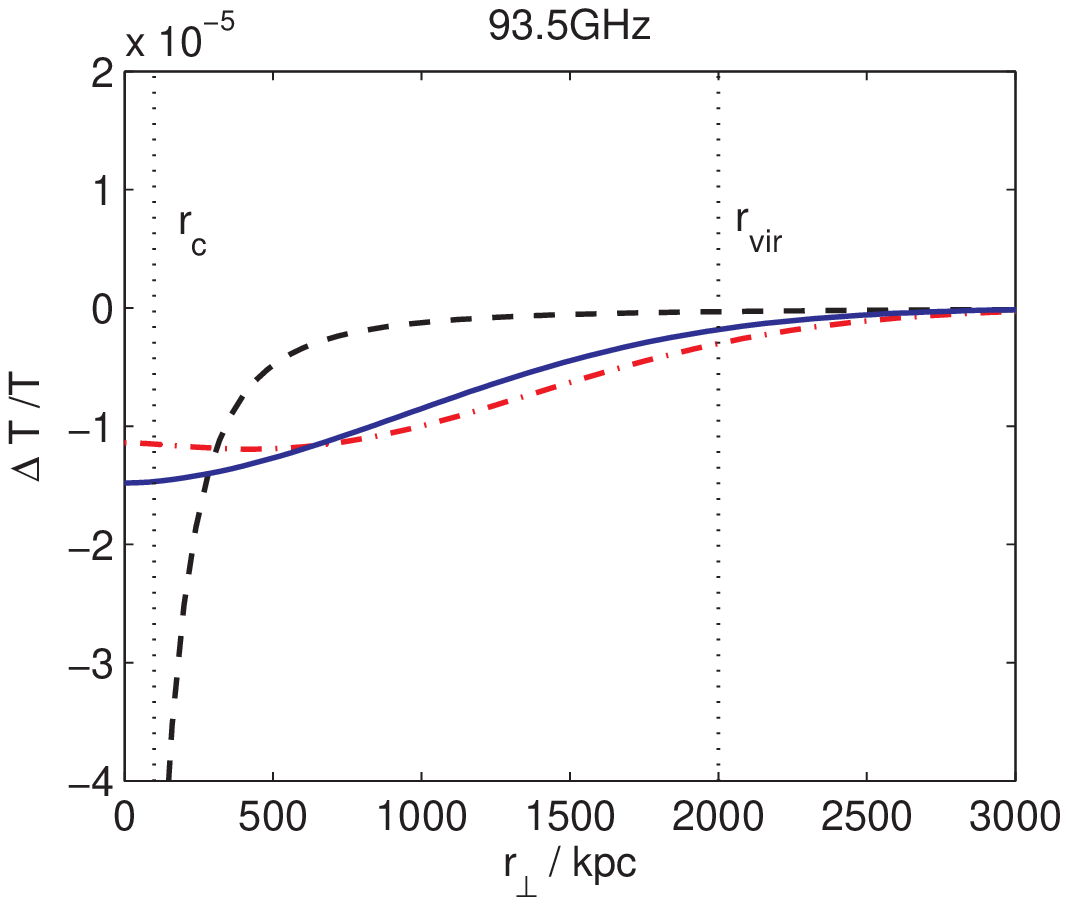}
\includegraphics[width = 5cm]{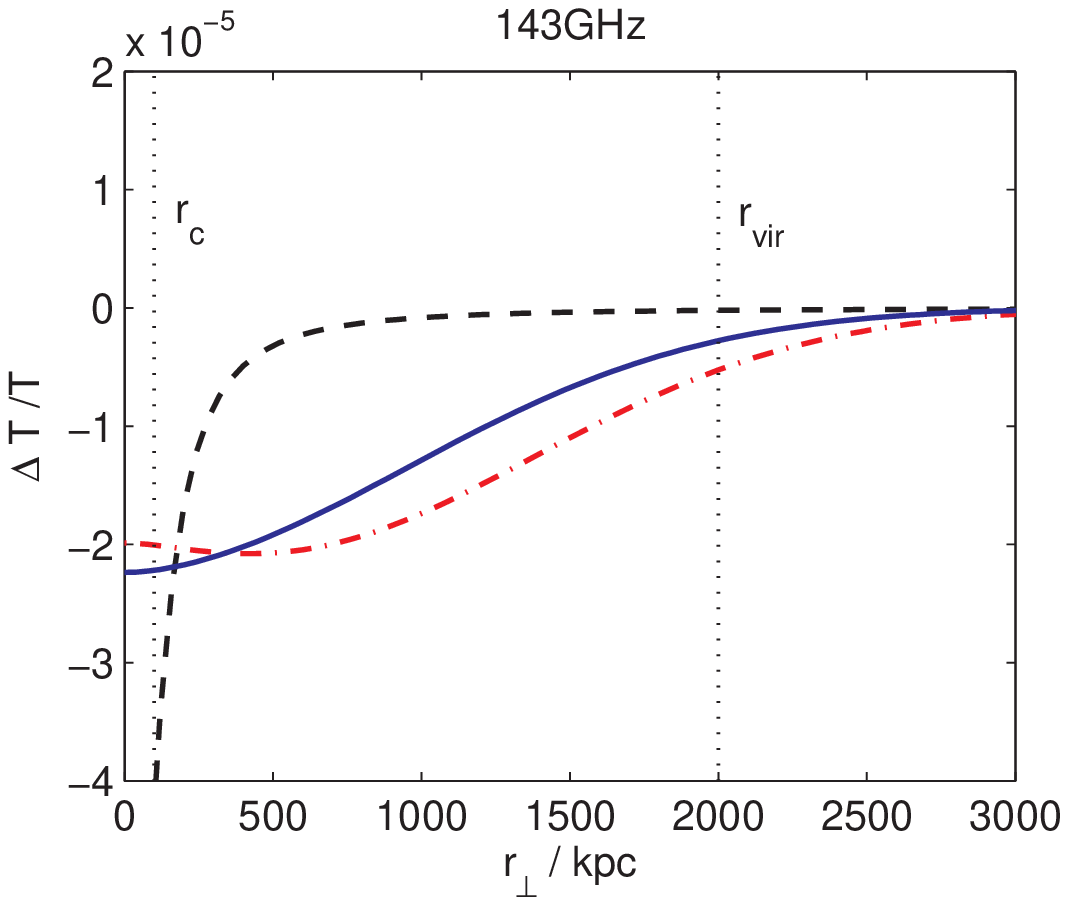}\includegraphics[width = 5cm]{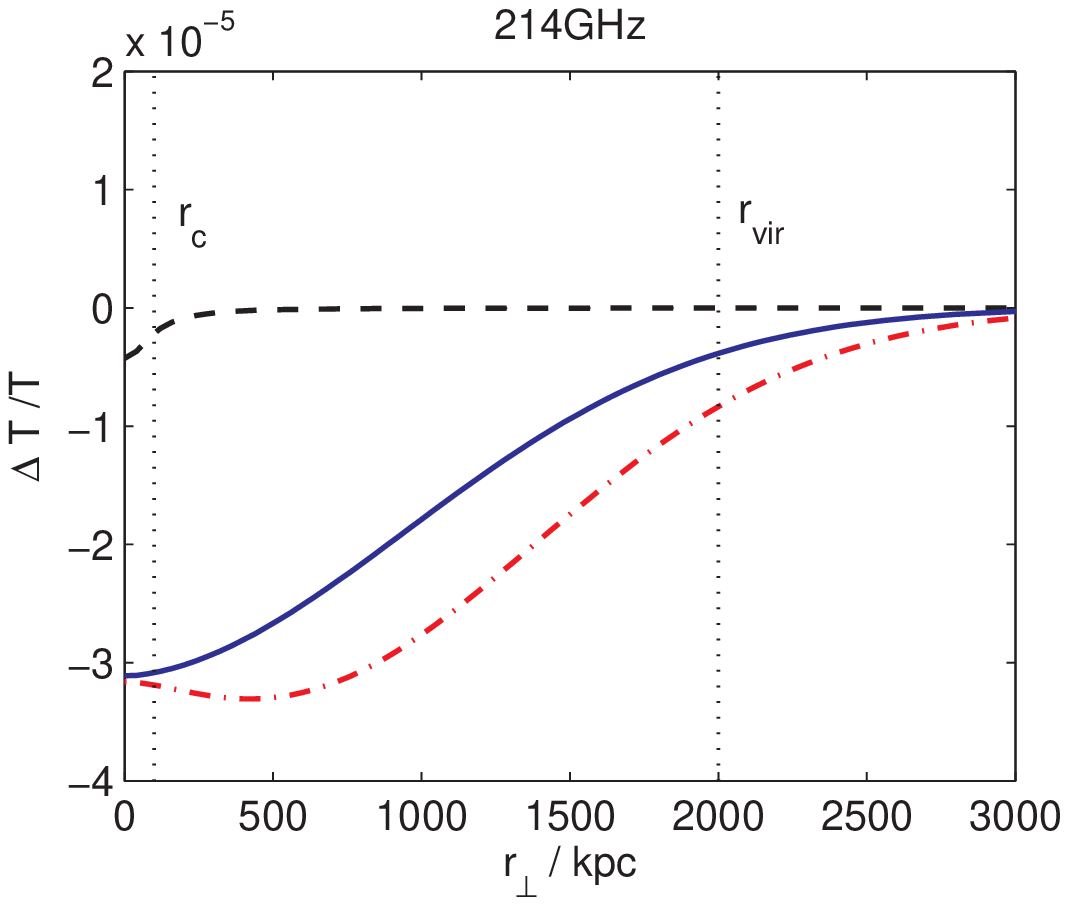}\includegraphics[width = 5cm]{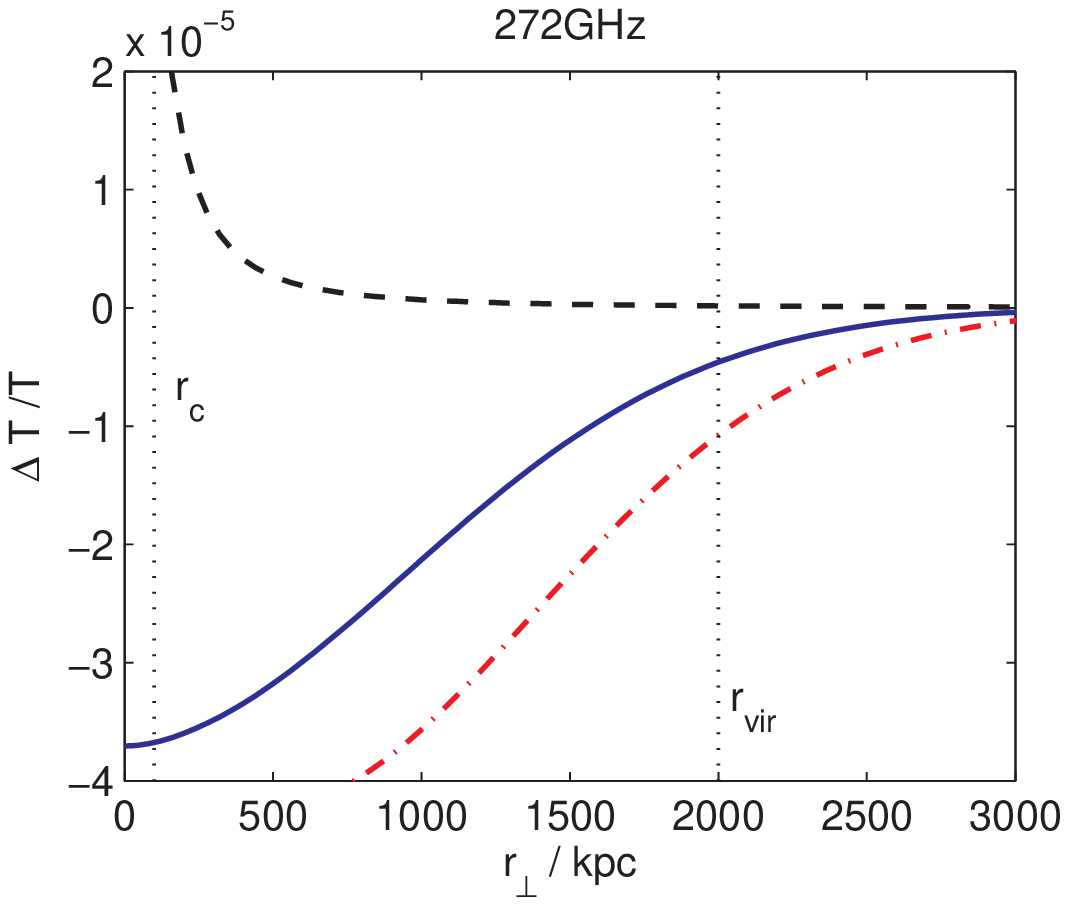}
\par\end{centering}

\caption{\label{fig:profile}Chameleonic (red dot-dashed line for $\alpha = -2$ and solid blue line for $\alpha = -5/3$) and thermal (dashed black line) SZ intensity decrement
profiles for a simulated cluster at six different frequencies (taken to be the same
as the frequency bands in the Coma cluster measurements). The simulated
cluster has $\Pphi(204\,{\rm  GHz}) = 3 \times 10^{-5}$ and $\tau_{0} = 5 \times 10^{-3}$ through the centre of the cluster, $r_{c}=100\,\mathrm{kpc}$ and $r_{\rm vir} = 2\,{\rm Mpc}
$. We assume $\eta=0.9$ and $\beta = \gamma = 1$.  }

\end{figure*}

For a constant $\left\langle \delta\mathbf{B}^2\right\rangle$, $\bar{n}_{\rm e}$ and $L_{\rm coh}$, we found that,
\be
\Pphi \propto L L_{\rm coh}^{1+\alpha_0} \left\langle \delta\mathbf{B}^2\right\rangle \bar{n}_{e}^{\alpha_0}, \nonumber
\ee
where $\alpha_{0} = \alpha$ if $-2 \leq \alpha < -1$ in the power spectrum model, and $\alpha_0 = -2$ in the cell model.  When  $\left\langle \delta\mathbf{B}^2\right\rangle$, $\bar{n}_{\rm e}$ and 
$L_{\rm coh}$ vary slowly along the line of sight we should change this to
\be
\Pphi \propto \int \dd z L_{\rm coh}^{1+\alpha_0}(z) \left\langle \delta\mathbf{B}^2(z)\right\rangle \bar{n}_{e}^{\alpha_0}(z) \dd z. \nonumber
\ee
We define $r_{\perp}$ to be the projected distance from the cluster centre for a given line of sight.  We also assume that the turbulent cluster magnetic field only extends out 
to some characteristic radius $r_{\rm max}$, and then dies off very quickly due to  $L_{\rm coh}$ growing very quickly or $\left\langle \delta\mathbf{B}^2(z)\right\rangle $ 
decreasing much faster than $\bar{n}_{\rm e}^{\eta}$.  We discuss reasonable values for $r_{\rm max}$ below.  With our assumed models for the radial dependence of the coherence length, magnetic field and electron density, we have
\be
\Pphi(r_{\perp}) \propto L_{\rm core}\left(1+\frac{r_{\perp}^2}{r_{\rm c}^2}\right)^{\frac{p}{2}} I_{x}(r_{\perp};p,r_{\rm max},r_{c}), \label{Pdrop}
\ee
where $p = 1+(1  +\alpha_0) \gamma - 3\beta(2\eta+\alpha_0)$, $L_{\rm core} = 2r_{\rm c}$ and
\be
I_{x} &=& \int_{0}^{\infty} \dd x \, \left(1+x^2\right)^{\frac{(p-1)}{2}} w(x;x_{\rm max}). \label{IxEqn}
\ee
We have introduced a window function, $w(x)$, to impose the rapid decay of magnetic turbulence for $r \gtrsim r_{\rm max}$. A simple model for $w(x;x_{\rm max})$ would be 
that $w = 1$ for $r^2 < r_{\rm max}^2$ and $0$ otherwise.  This corresponds to $w = H(x_{\rm max}-x)$ where $H$ is the heaviside function, and
$$
x_{\rm max} = \sqrt{\frac{r_{\rm max}^2-r_{\perp}^2}{r_{\rm c}^2+r_{\perp}^2}}.
$$
Alternatively, we could impose a smoother decay of the turbulent magnetic field by taking $w = \exp(-r^2/r_{\rm max}^2)$, which corresponds to 
\be
w(x) = \exp\left(-\frac{r_{\perp}^2 +(r_{c}^2+r_{\perp}^2)x^2}{r_{\rm max}^2}\right). \label{wEqn}
\ee
In what follows we approximate $w(x)$ by this latter form. 

\begin{figure*}[htb!]
\begin{centering}
\includegraphics[width = 7.5cm]{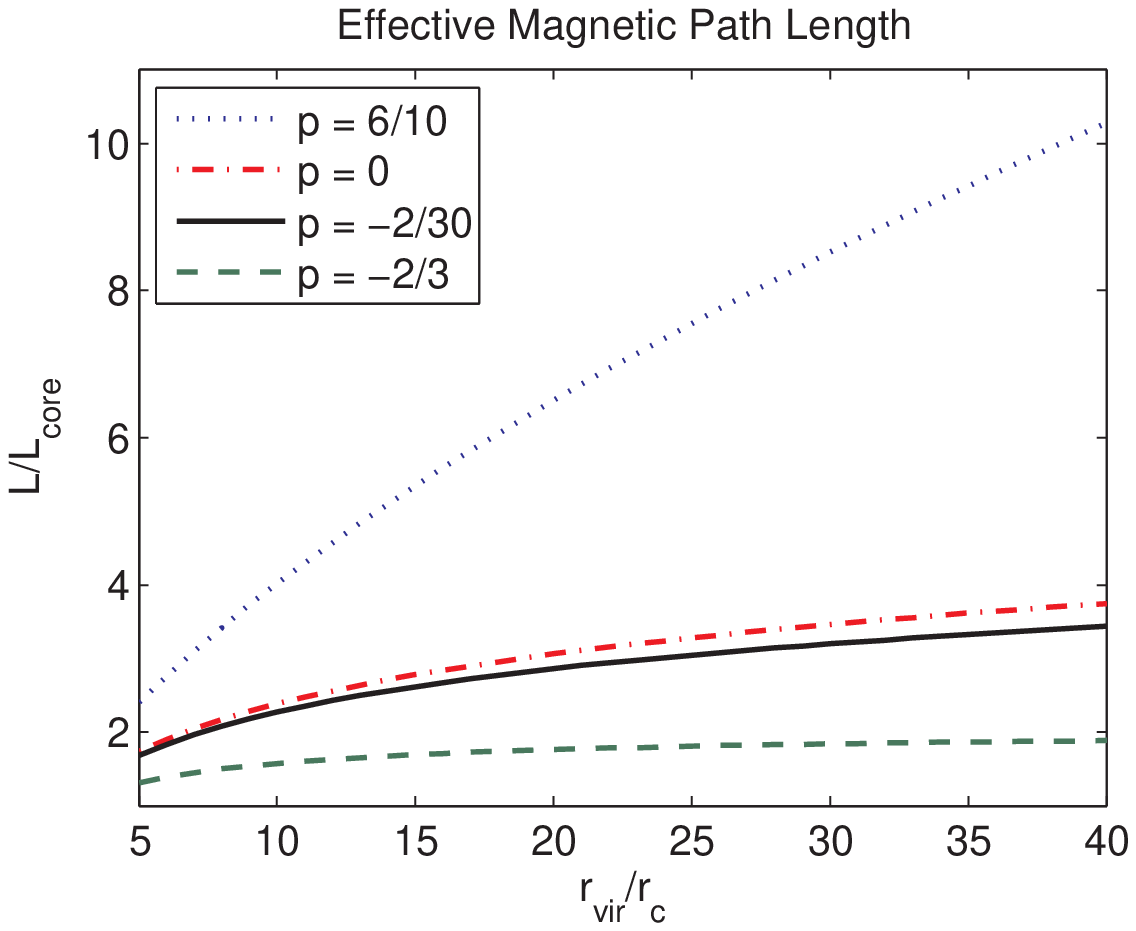}\includegraphics[width = 7.5cm]{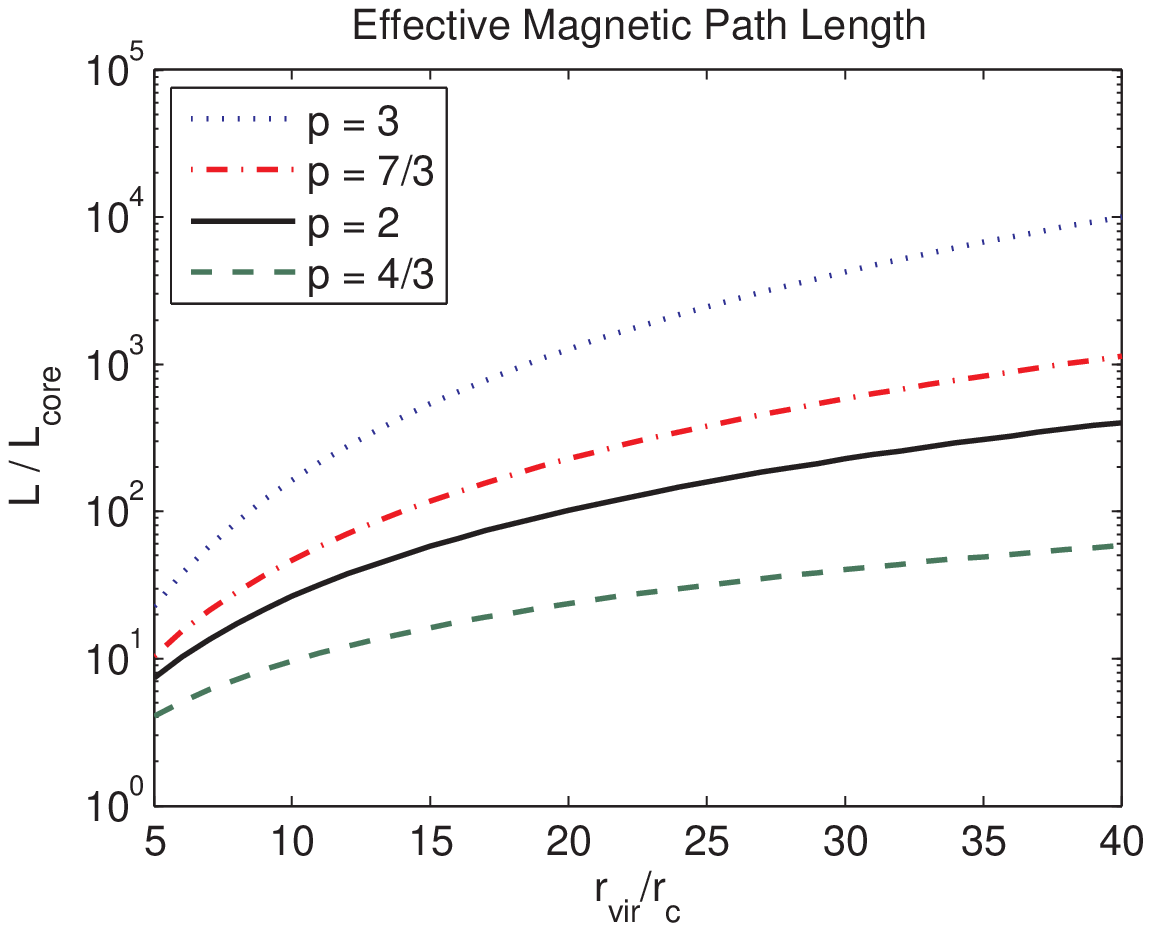}
\par\end{centering}

\caption{\label{Fig:Path}The effective path length $L$  through the magnetic region.  The leftmost plot shows $p$ values corresponding to $\eta = 0.9$ and $\eta = 1$ in the 
Kolmogorov ($\alpha = -5/3$) and cell ($\alpha  = -2$) magnetic field models when $\beta = \gamma  =1$.  These are the values of $\eta$ suggested by observations and simulations.
 The rightmost plot shows $p$ values corresponding to $\eta = 1/2$ and $\eta = 2/3$ again in the Kolmogorov and cell magnetic field models with
  $\beta = \gamma  =1$; these are the values of $\eta$ suggested theoretically. In the Kolmogorov  model $\eta = \left\lbrace 1/2, 2/3, 0.9, 1\right\rbrace$ corresponds to $p = \left\lbrace 7/3, 4/3,-2/30, -2/3\right\rbrace$. In the cell  model the same values of $\eta$ give $p = \left\lbrace 3, 2,6/10, 0\right\rbrace$. }

\end{figure*}

All that remains is to determine a reasonable estimate for the largest radius to which the turbulent 
cluster magnetic field extends, $r_{\rm max}$. We note that whenever $p>0$, the dominant contribution to the integral in Eq. (\ref{Pdrop}) will come from the largest values of 
$x$, i.e. from $r \sim r_{\rm max}$.  The value of $r_{\rm max}$ 
will, therefore, play an important role in such scenarios. Assuming $\beta = \gamma = 1$, in the cell magnetic field model ($\alpha_0 = -2$) we have $p_{\rm cell} = 6(1-\eta)$ and 
so $p_{\rm cell} > 0$ for $\eta < 1$, while if we assume a Kolmogorov spectrum with $\alpha_0 = -5/3$ we have $p_{\rm Kolmo} = 6(8/9-\eta)$ and so $p_{\rm cell} > 0$ for 
$\eta < 8/9$. In Ref. \cite{Ryu08}, the quantity $\varpi_{\rm rms}t_{\rm age}$ is introduced where $t_{\rm age}$ is the age of the Universe and $\varpi_{\rm rms}$ is the root mean squared 
value of the vorticity; $\varpi = \nabla \times \mathbf{v}$, where $\mathbf{v}$ is the peculiar velocity.  In Ref. \cite{Ryu08} it is estimated that when 
$\varpi_{\rm rms}t_{\rm age}$ is greater than $\mathcal{O}(1)$, turbulence has had time to develop.  In a virialized structure we have the following order of magnitude estimates: 
$\varpi \sim O(v(r)/r)$ and $v^2(r) \sim O(M(r)/r)$ where $M(r)$ is the mass inside a radius $r$.   Thus,
$$
\varpi_{\rm rms}t_{\rm age} \sim \Oo(\sqrt{\bar{\rho}(r)/\rho_{\rm c}}),
$$
where $\bar{\rho}(r)$ is the average density inside a radius $r$ and $\rho_{\rm c}$ is the cosmological critical density today. Beyond the virial radius, we expect that 
$\varpi^2 < \Oo(M(r)/r^2)$, as the vorticity acts to slow down the collapse.  Roughly then we expect $\omega_{\rm rms}t_{\rm age}$ to be smaller than a few when 
$r \gg r_{\rm vir}$, and larger at smaller radii.  We choose $r_{\rm max}$ so that the chameleonic SZ effect dies off quickly for $r > r_{\rm vir}$. 
With the choice Gaussian window function given in Eq. (\ref{wEqn}), a suitable choice is $r_{\rm max} \approx r_{\rm vir}$.

We have the following approximate scalings for the thermal and chameleonic SZ effects:
\be
\frac{\Delta T_{\rm SZ}}{T_{0}}&\propto& \left(1+\frac{r_{\perp}^2}{r_{\rm c}^2}\right)^{\frac{1}{2}-\frac{3\beta}{2}},\nonumber \\
\frac{\Delta T_{\rm CSZ}}{T_{0}} &\propto& \left(1+\frac{r_{\perp}^2}{r_{\rm c}^2}\right)^{\frac{p}{2}} I_{x}(r_{\perp};p, r_{\rm vir},r_{\rm c}), \nonumber
\ee
where $I_{x}$ is given by Eq. (\ref{IxEqn}) and we take the Gaussian window function given in Eq. (\ref{wEqn}) to impose the decay of the turbulent magnetic field for 
$r > r_{\rm max} \approx r_{\rm vir}$. For $r_{\perp} \ll r_{\rm vir}$, the chameleonic contribution to the SZ effect will decrease less rapidly than the thermal one when,
$$
p+3\beta-1 = (1+\alpha_0)\gamma + 3(1-\alpha_0-2\eta) \beta > 0.
$$
For $\gamma \approx \beta$, and $\eta \leq 1$ this holds for all $-2 \leq \alpha_0 < -1$. Hence, even if the thermal SZ effect dominates near the centre of the cluster, measurements of the SZ effect at radii $r_{\rm c} \ll r \ll r_{\rm vir}$ might be able to detect the chameleonic effect.

In Fig. \ref{fig:profile} we plot the predicted profiles for the
chameleonic and thermal SZ effects at different frequencies for a model cluster, with $\eta = 0.9$ and $\beta=\gamma = 1$. We assume 
that the cluster properties are similar to those of the Coma cluster and assume $\Pphi = 3\times 10^{-5} $ at $\nu = 204\,{\rm GHz}$ for a line of sight through the centre of the
cluster. This is roughly the 1$\sigma$ confidence limit derived earlier from Coma cluster observations.  We take $r_{\rm c} \approx 100\,{\rm kpc}$, and $r_{\rm max} \approx r_{\rm vir} \approx 2\,{\rm Mpc}$. The optical depth for a line of sight through the centre of cluster is taken to be $\tau_{0} = 5\times 10^{-3}$.    The 
black dashed line shows the usual thermal SZ effect, the dot-dashed red line is the chameleonic SZ effect for $\alpha_0 = -2$ (cell model) and the solid blue line is the 
chameleonic SZ effect for $\alpha_0 = -5/3$ (Kolmogorov turbulence model). In general, the smaller the value of $\eta$, the more slowly the chameleonic SZ effect decays with distance
 from the cluster centre.  We note that for $30\,{\rm GHz}\lesssim \nu \lesssim 300\,\rm  GHz$, a chameleonic SZ effect of this magnitude dominates over the thermal SZ effect for
  $r_{\perp} \gtrsim 500\kpc$. At frequencies around $150\,{\rm GHz} \lesssim \nu \lesssim 275\,{\rm GHz}$, the chameleonic SZ effect is dominant for $r_{\perp} \gtrsim r_{\rm c}$. Measurements of the CMB intensity decrement associated with clusters over such a frequency range and for $r \gtrsim r_{\rm c}$ could be particularly useful in constraining the properties of chameleon-like theories.

There has been some suggestion of a discrepancy between X-ray emission data as measured by ROSAT, which determines the electron density distribution,
and the SZ intensity decrement as measured by WMAP \cite{ROSATvsWMAP}. In the analysis the expected signal derived from X-ray measurements is fitted to the observed SZ effect at large radii. They conclude that the
observed intensity decrement at the centre of the cluster is then less than the predicted SZ effect. However a possible alternative is the existence of a chameleon-like scalar field, which would cause an enhancement in the SZ effect at large radii. The WMAP data includes measurements at $94\,{\rm GHz}$ for which the chameleon effect could be significant. 

In \S \ref{sub:Mag}, we estimated $\Pphi$ in terms of the properties of the cluster magnetic fields, and approximated the path length, $L$, through the magnetic region for a light
 ray passing through the cluster as being equal to twice the core radius, i.e. $L \approx L_{\rm core} = 2r_{\rm c}$. However, the relatively flat radial profile (for 
 $r \lesssim r_{\rm vir}$) of the chameleonic SZ effect derived above suggests that $L$ is generally much larger than $L_{\rm core}$, and for $p > 0$ is better approximated as being $\Oo((r_{\rm vir}/r_{\rm c})^{p}L_{\rm core})$.  Using the model introduced above for the radial behaviour of the electron density, the magnetic field and  $L_{\rm coh}$, we find that the effective path length is given by:
\be
L = L_{\rm core} I_{x}(0;p,r_{\rm max}\approx r_{\rm vir},r_{c}). \nonumber
\ee
It follows from Eq. (\ref{IxEqn}) and Eq. (\ref{wEqn}) that $L/L_{\rm core}$ depends on two parameters $p$ and $r_{\rm vir}/r_{\rm c}$.  In Fig. \ref{Fig:Path}, we plot 
$L/L_{\rm core}$ against $r_{\rm vir}/r_{c}$ for different values of $p$. Reasonable estimates for $\eta$ are $\eta = 1/2$, 
$2/3$, $0.9$ and $1$. $\eta = 1/2$ and $\eta = 2/3$ are based upon theoretical expectations, the former being for equipartition of magnetic and thermal energy and the latter 
being if the magnetic field is frozen in.  Observations and simulations suggest that these values are unrealistic and instead indicate $\eta = 0.9$ and $\eta = 1$ respectively. 
With $\beta = \gamma = 1$ and $\eta = \{0.9,1\}$, $p = \{-2/30,-2/3\}$ in the 
Kolmogorov model, and $p = \{6/10,0\}$ in the cell model. We plot these values of $p$ for $5 < r_{\rm vir}/r_{\rm c} < 40$ in the 
leftmost plot of Fig. \ref{Fig:Path}. We can see that in all four cases $L_{\rm core} \lesssim L \lesssim 11 L_{\rm core}$ for the range of $r_{\rm vir}/r_{\rm c}$ considered. 
Since $\Pphi \propto g_{\rm eff}^2$, this corresponds to no more than a factor $3$ or so enhancement in sensitivity to $g_{\rm eff}$.  
If instead we consider $ \eta = \{1/2,2/3\}$ then $p=\{7/3,4/3\}$ in the Kolmogorov model and $p = \{3,2\}$ in the cell model. We plot these values of
 $p$ in the rightmost plot of  Fig. \ref{Fig:Path}.  In these cases, we see that $L \gg L_{\rm core}$, and as expected $
 L/L_{\rm core} \sim \Oo( (r_{\rm vir}/r_{\rm c})^{p})$.  Since $\eta =1/2$ and $\eta=2/3$ do not appear to be supported by observations, and because the structure of cluster magnetic fields at radii $r \gg r_{\rm core}$ is far from 
 certain, we do not view values of $L \gg L_{\rm core}$ as being particularly realistic. The more realistic scenarios are those shown in the leftmost plot in Fig. \ref{Fig:Path} where $1 \lesssim L/L_{\rm core} \lesssim 11$.

\section{Modifications to the CMB Polarization\label{sec:Modification-Polarisation}}
In addition to modifying the intensity of light that has passed through magnetic regions, we found in \S \ref{sec:Optics} that the presence of a chameleon-like scalar field also 
alters the polarization of the light. We define $Q_{0}$,  $U_{0}$ and $V_{0}$ to be the fractional intrinsic CMB polarizations. We found that after mixing with the chameleon in a
 magnetic region,
\be
Q &=& Q_{0} + \mathcal{Q}_{\rm q}, \nonumber \\
U &=& U_{0} + \mathcal{Q}_{\rm u}, \nonumber \\
V &=& V_{0} + \mathcal{Q}_{\rm v}. \nonumber
\ee
We use angled brackets to denote an average over many lines of sight at fixed frequency, i.e. fixed $\Delta$. In both the cell and power spectrum models we find,
\be
\left\langle \mathcal{Q}_{\rm q} \right\rangle &=& \left\langle Q-Q_{0}\right\rangle = \mathcal{Q}_{0} \cos 2\theta_{\rm reg}, \nonumber \\
\left\langle \mathcal{Q}_{\rm u} \right\rangle &=& \left\langle U-U_{0}\right\rangle = \mathcal{Q}_{0} \sin 2\theta_{\rm reg}, \nonumber \\
\left\langle \mathcal{Q}_{\rm v} \right\rangle &=& \left\langle V-V_{0}\right\rangle = 0,\nonumber 
\ee
where $B^{\rm reg}_{x} = B^{\rm reg}\cos \theta_{\rm reg}$ and $B^{\rm reg}_{y} = B^{\rm reg}\sin \theta_{\rm reg}$ define $\theta_{\rm reg}$. Detailed derivations of these 
formulae are presented in Appendix A. In the cell model, 
$$\mathcal{Q}_{0} \simeq \frac{1}{2}\left(\frac{2B_{\rm reg}\omega}{M_{\rm eff}m_{\rm eff}^2}\right)^2 \sin^2 \bar{\Delta}, $$
whereas in the power spectrum model,
\be
\mathcal{Q}_{0} &\simeq& \frac{1}{4}\left(\frac{2B_{\rm reg} \omega}{M_{\rm eff} \bar{m}_{\rm eff}^2}\right)^2 I_{\rm N}^3 \nonumber  \\ 
&&- \frac{1}{4}\left(\frac{2B_{\rm reg} \omega}{M_{\rm eff} \bar{m}_{\rm eff}^2}\right)^2 \cos \left(\frac{\bar{m}_{\rm eff}^2 L}{2\omega}\right) \nonumber \\ 
&& + \frac{B_{\rm reg}^2 L }{16M_{\rm eff}^2 \bar{n}_{\rm e}^2} I_{\rm N}^2 W_{\rm N}(k_{\rm crit}). \nonumber
\ee
In both cases $\mathcal{Q}_{0} = \mathcal{P}^{\rm reg}$ where $\mathcal{P}^{\rm reg}$ is the contribution of the regular magnetic field to the conversion probability.  In addition
 to the contribution from mixing in the regular magnetic field along a given line of sight, the Stokes parameters will also pick up a contribution from the turbulent magnetic 
 field.  The magnitude and direction of this contribution is determined by the direction and size of the magnetic field fluctuations along the line of sight, and so vanishes when
  averaged over many lines of sight. Along any given line of sight it would be observed as an essentially random contribution, and would be present in the variance of the 
  $\mathcal{Q}_i$, given by $\sigma_{i}^2(\bar{\Delta}, \bar{\Delta})$:
\be
\sigma_{i}^2(\Delta_1,\Delta_2) = \left\langle \mathcal{Q}_{i}(\Delta_1)\mathcal{Q}_{i}(\Delta_2)\right\rangle - \left\langle \mathcal{Q}_{i}(\Delta_1)\right\rangle\left\langle\mathcal{Q}_{i}(\Delta_2)\right \rangle. \nonumber
\ee
A detailed calculation of these $\sigma_{i}$ is given in Appendix A. We find similar results for both the cell and power spectrum models.  In the latter we have,
\be
\sigma_{\rm q}^2(\bar{\Delta},\bar{\Delta}) &\approx& \frac{1}{2}\bar{\mathcal{P}}^2_{\gamma \leftrightarrow \phi} + \frac{1}{2}\mathcal{Q}_{0}^2 \cos 4\theta_{\rm reg}, \nonumber \\
\sigma_{\rm u}^2(\bar{\Delta},\bar{\Delta}) &\approx& \frac{1}{2}\bar{\mathcal{P}}^2_{\gamma \leftrightarrow \phi} - \frac{1}{2}\mathcal{Q}_{0}^2 \cos 4\theta_{\rm reg}, \nonumber \\
\sigma_{\rm v}^2(\bar{\Delta},\bar{\Delta}) &\approx& \frac{1}{2}\bar{\mathcal{P}}^2_{\gamma \leftrightarrow \phi} - \frac{1}{2}\mathcal{Q}_{0}^2 \,.\nonumber 
\ee
Hence, looking along a single line of sight at a single frequency, we would expect $\vert Q\vert$, $\vert U\vert$, 
$\vert V\vert \sim \mathcal{O}(\bar{\mathcal{P}}_{\gamma \leftrightarrow \phi}/\sqrt{2})$, since $ \bar{\mathcal{P}}_{\gamma \leftrightarrow \phi}$ is dominated by the 
contribution from the random magnetic field ($ \bar{\mathcal{P}}_{\gamma \leftrightarrow \phi}\gg \mathcal{P}^{\rm reg}=\mathcal{Q}_0 $). The best upper bound on 
$\bar{\mathcal{P}}_{\gamma \leftrightarrow \phi}$ from the Coma cluster SZ measurements was,
$$
\bar{\mathcal{P}}_{\gamma \leftrightarrow \phi}(204\,{\rm GHz}) < 6.2 \times 10^{-5},
$$
along lines of sight passing though the Coma cluster core.  Along such lines of sight, we might expect chameleonic contributions to $Q$, $U$ and $V$ that are as large as a 
${\rm few} \times 10^{-5}$. If this were the case then the magnitude of all three Stokes parameters would be roughly $10-100\,\mu{\rm K}$, which is orders of magnitude larger than
 the measured $0.1-1\mu{\rm K}$ level of $Q$ and $U$ and the even lower bound on the circular polarization, $V$. One might imagine, therefore, that CMB polarization measurements 
 could greatly improve the upper limit on $\bar{\mathcal{P}}_{\gamma \leftrightarrow \phi}$, say placing it around the $10^{-7}$ level.  We found in \S \ref{sub:ComaCluster} that
  in the Kolmogorov model for the power spectrum of magnetic fluctuations, $\mathcal{P}_{\rm Coma}^{\rm (Kolmo)}$ could be almost as large as $10^{-6}$ with $g_{10} \sim O(1)$.  An upper limit of $10^{-7}$ could therefore constrain $M_{\rm eff} \gtrsim 10^{10}\,{\rm GeV}$. However, there are two important caveats of which we must take account and which limit the ability of realistic experiments to detect the chameleonic contribution to the CMB polarization.

Firstly, the random effect (whose variance is given by the $\sigma_{i}^2$) vanishes when one averages over many lines of sight which have taken different paths through the random
 magnetic regions.  The coherence length, $L_{\rm coh}$, provides the typical length scale of the random magnetic regions. We therefore expect that roughly parallel light rays 
 with a perpendicular separation $d \gg L_{\rm coh}$ will have taken different paths through the magnetic regions.  If the cluster in question is a distance, $D_{\rm clust} \gg
 d$, from us then light rays with a separation, $d \lesssim L_{\rm coh}$, will subtend an angle $\lesssim \theta_{\rm coh}$ when observed on the sky, where 
$$
\theta_{\rm coh} = \frac{10800}{\pi} \frac{L_{\rm coh}}{D_{\rm clust}}\, {\rm arcmin}.
$$
As an example we consider the values for these quantities for the Coma and Hydra A clusters: $(D_{\rm clust},L_{\rm coh}) \sim (100\,{\rm Mpc}, 1\,{\rm kpc})$ and 
$\sim (50\,{\rm Mpc}, 4\,{\rm kpc})$ respectively. This results in $\theta_{\rm coh} \sim 0.03\,{\rm arcmin}$ for Coma and $\theta_{\rm coh} \sim 0.27 {\rm arcmin}$ for Hydra A. 
 In order to resolve the random chameleonic contribution to the CMB polarization, a survey must have an angular resolution better than $\theta_{\rm coh}$.  The polarization of 
 the CMB in the direction of galaxy clusters has
not so far been measured, but there are plans to incorporate a polarization
detector at the South Pole Telescope, SPT-pol, which is expected to
be completed in the next few years \cite{SPTelescope}. SPT is searching
for clusters on the sky by exploiting the SZ effect. Each pixel of detectors
at SPT represents one arc-minute, and the aim is to have $10\,\mu{\rm K}$ sensitivity per pixel.  For a cluster such as Coma where $L_{\rm coh}$ is relatively small, this one arc-minute angular resolution is not enough to resolve the random chameleonic  contribution to the polarization. Even for a cluster such as Hydra A where $L_{\rm coh}/D_{\rm clust}$ is larger, it is still a factor of four larger than $\theta_{\rm coh}$.   In general, resolving the random chameleonic contribution to the Stokes parameters will require an ability to resolve the polarization signal on scales smaller than a few arc-seconds.

Even if one has the required angular resolution, there is still an additional technical limit on the ability to detect the random chameleonic modification to the Stokes parameters.  The expressions given for the $\sigma_{i}^2$ assume that measurements are made at precisely one frequency. Any realistic measurement will, however, not be entirely monochromatic but represent an average over some frequency band.  Suppose one makes a measurement at a nominal wavelength, $\bar{\lambda}$, which is actually an average (with some window function) over a range $\lambda-\bar{\lambda} \in (-\delta \lambda, \delta \lambda)$.  This corresponds to the value of $\bar{\Delta}$ running between $\bar{\Delta}-\delta\Delta$ and $\bar{\Delta}+\delta\Delta$  where $\delta \Delta = \bar{\Delta} (\delta \lambda/\lambda)$.  The measured variance in the Stokes parameters will then be given by,
\be
\hat{\sigma}^2_{i}(\bar{\Delta}) = \frac{1}{(2\delta \Delta)^2}\int_{\bar{\Delta}-\delta \Delta}^{\bar{\Delta}+\delta \Delta} \dd \Delta_{1} \int_{\bar{\Delta}-\delta \Delta}^{\bar{\Delta}+\delta \Delta} \dd \Delta_{2}\, \sigma_{i}^2(\Delta_1,\Delta_2). \nonumber
\ee
A full calculation of the $\sigma^2_{i}(\Delta_1, \Delta_2)$ shows that the dominant term in it drops off as $1/(\Delta_1-\Delta_2)$ when $\vert \Delta_1-\Delta_2\vert \gg 1$.  It 
follows that when $\delta \Delta \gg 1$, $\hat{\sigma}^2_{i}$ is suppressed by a factor $\mathcal{O}(1/\delta \Delta^2)$ relative to its value when $\delta \Delta = 0$.  Now,
\be
\delta \Delta \approx (4.3\times 10^{3})\left(\frac{n_{\rm e}}{10^{-3}\,{\rm cm}^{-3}}\right) \left(\frac{L}{100\,{\rm kpc}}\right) 
\left(\frac{\delta \lambda}{1\,\mu{\rm m}}\right). \nonumber
\ee
Typically for CMB radiation $\lambda \sim 1\,{\rm mm} \text{--} 1\,{\rm cm}$, and so to achieve $\delta\Delta \lesssim 1$, we would require $\delta \lambda/\lambda < 10^{-6}-10^{-7}$ which is many orders of magnitude smaller than the typical spectral resolution of CMB experiments, $\delta \lambda/\lambda \sim 10^{-4}-10^{-2}$.

It is clear then that both the angular and spectral resolution of detectors would have to improve significantly before the random chameleonic contribution to the CMB polarization
 could be detected.  The smaller regular component,  however, does not suffer from such issues, although unfortunately it does not source the (more unusual) circular polarization
  in the CMB. In the cell model the contribution from the regular magnetic field is a factor of $1/N$ smaller than $\bar{\mathcal{P}}_{\gamma \leftrightarrow \phi}$.    In the 
  power spectrum model, however, electron density fluctuations result in some enhancement of the regular magnetic field conversion probability.  The regular contributions to both
   $Q$ and $U$ are generally dominated by the term in $\mathcal{Q}_{0}$  coming from the electron density fluctuations:
\be
\mathcal{Q}_{0} &\approx& \frac{B_{\rm reg}^2 L }{16M_{\rm eff}^2 \bar{n}_{\rm e}^2} I_{\rm N}^2 W_{\rm N}(k_{\rm crit}). \nonumber
\ee
Assuming that $k^2 P_{\rm N} = 2(2\pi)^{1/3}C_{\rm N}^2 k^{\alpha}$ for $k \gtrsim L_{\rm coh}^{-1}$, we found in \S \ref{sub:PowerSpectra} that for a Kolmogorov spectrum with 
$\alpha = -5/3$,
\be
2(2\pi)^{1/3}C_{\rm N}^2 &\approx& (0.27\text{--}0.45)\bar{n}_{\rm e}^2 (I_{\rm N}-1) L_{\rm coh}^{-2/3}.\nonumber
\ee
Hence in the Kolmogorov model,
\be 
\mathcal{Q}_{0}  &\sim& 10^{-2} \frac{I_{\rm N}^2(I_{\rm N}-1) B_{\rm reg}^2 L L_{\rm coh}^{-2/3} }{M_{\rm eff}^2 } k_{\rm crit}^{-5/3}, \nonumber \\
&\approx& (0\text{--}1.5 \times 10^{-9}) g_{10}^2 \nu_{214}^{5/3}B_{1 \mu{\rm G}}^2 L_{200} \nonumber \\
& & n_{0.01}^{-5/3} \left(L_{1\kpc}^{\rm coh}\right)^{-2/3}, \nonumber
\ee
where as before $I_{\rm N} = 1\text{--}2$, $g_{10} = 10^{10}\GeV/M_{\rm eff}$, etc. and we define $B_{1 \mu{\rm G}}=B_{\rm reg}/1\mu {\rm G}$. For the Coma cluster 
with $L\approx 198\,\mathrm{kpc}$, $\bar{n}_{\rm e}\approx 4\times 10^{-3}\,{\rm cm}^{-3}$, $B_{\rm reg} \approx 0.2\,\mu{\rm G}$ and $L_{\rm coh} \approx 1\,{\rm kpc}$ this
gives,
\be
\mathcal{Q}_{0}^{\rm Coma} \sim (0\text{--}2.9)\times 10^{-10} g_{10}^2\nu_{214}^{5/3}, \nonumber
\ee
whereas for Hydra A with $L \approx 260\,{\rm kpc}$, $\bar{n}_{\rm e} \approx 10^{-2}\,{\rm cm}^{-3}$, $L_{\rm coh} \approx 4\,{\rm kpc}$ and 
$B_{\rm reg} \approx 6\,\mu{\rm G}$ we have,
\be
\mathcal{Q}_{0}^{\rm Hydra} \sim (0\text{--}3.0) \times 10^{-9} g_{10}^2 \nu_{214}^{5/3}. \nonumber
\ee
For the chameleon-induced linear polarization to be at a similar level to the intrinsic CMB polarization one would require $\mathcal{Q}_{0} \sim {\rm few} 
\times 10^{-7}$.  For the Coma cluster this would require $g_{10} \gtrsim 27$ corresponding to $\mathcal{P}_{\rm Coma}(214\,{\rm GHz}) \sim {\rm few} \times 10^{-4}$, which is
strongly ruled out by our analysis of the SZ effect.  However in Hydra A, where the large scale magnetic field is much stronger, one could achieve $\mathcal{Q}_{0} \sim
{\rm few} \times 10^{-7}$ with only $g_{10} \approx 2.6$ which is a factor of 4 smaller than the lowest upper bound on $g_{10}$ found previously from the Coma SZ 
measurements.  It is feasible, therefore, that future searches for CMB polarization in the direction of galaxy clusters could, in some cases, reveal an enhancement
of polarization at higher frequencies. However this would only be expected to occur for clusters, such as Hydra A, with suitably large regular magnetic
fields, $\mathcal{O}(10\mu{\rm G})$.

\section{Conclusions\label{sec:Conclusions}}

The chameleon was first introduced to explain the possible scenario of a very light scalar field existing in the cosmos with a gravitational strength coupling to matter and yet 
not being detected on Earth as a fifth-force. The theory has since been extended to allow supergravitational strength couplings and different forms of the matter
coupling. Recently, there has been interest in predicting the astrophysical
implications of a direct coupling between chameleon-like fields and the photon
\cite{Brax07, BurrageSN, Burrage08}. This analysis
applies not only to the chameleon model but also other chameleon-like theories such as the Olive--Pospelov model designed to explain a variation in the fine-structure constant,
$\alpha_{\rm em}$.  

In this paper we have specifically considered the effect of a chameleon-like scalar field on the CMB intensity and polarization in the direction of galaxy clusters. The chameleon
 mixes with CMB photons in the magnetic
field of the intracluster medium. The resulting modification to the CMB intensity could be observed as a correction to the existing Sunyaev--Zel'dovich effect. Measurements of the CMB 
polarization in the direction of clusters are still waiting to be made, but the chameleon effect could potentially be significant in the polarization foregrounds as well.

Predictions of the chameleon effect are strongly dependent on the assumed structure of the intracluster magnetic field. In this work we have considered two different models for
the magnetic structure: a simplistic cell model in which the cluster properties are uniform but the magnetic field has a random orientation in each of the $N$ cells along the 
light path; and a potentially more realisistic power spectrum model in which fluctuations in the magnetic field and electron density are allowed over a range of length scales, 
with the small-scale power spectra taken to correspond to three-dimensional Kolmogorov turbulence.

The conversion probability, $\Pbar$, in the cell model for magnetic fluctuations is given by,
\be
\mathcal{P}^{(\rm cell)}(L,\nu)  &=& \left(3.2 \times 10^{-10}\right)g_{10}^2 B_{10\mu {\rm G}}^2 n_{0.01}^{-2}\nonumber\\ && L_{200} (L^{\rm coh}_{1\kpc})^{-1} \nu_{214}^2,\nonumber
\ee
while in the Kolmogorov turbulence model it is predicted to be slightly larger with 
\be
&\mathcal{P}^{(\rm  Kolmo)}(L,\nu) = \left(2.4\text{--}3.8 \times 10^{-7}\right)g_{10}^2B_{10\mu {\rm G}}^2 n_{0.01}^{-5/3} & \nonumber \\ &\nu_{214}^{5/3}L_{200}\left(L^{\rm coh}
_{1 \kpc}\right)^{-2/3} \left(\frac{I_{\rm N}^{2}(2I_{\rm N}-1)}{12}\right). &\nonumber
\ee
The resulting change to the CMB intensity contributing to the SZ effect is given by,
\be
\frac{\Delta I_{\rm CSZ}}{I_{0}} = -\Pbar(L). \nonumber
\ee

The strong dependence of the conversion probability on frequency allows us to compare measurements of the SZ effect across a range of frequencies, with our predicitions for the combined thermal SZ and chameleonic SZ effects. A maximum likelihood analysis of the SZ observations of the nearby Coma cluster leads to a constraint on the chameleon-photon conversion probability:
\be
{\mathcal{P}}_{\rm Coma}(204\,{\rm GHz}) < 6.2 \times 10^{-5}\quad (95\% {\rm CL}). \nonumber
\ee
This can be converted into a constraint on the chameleon-photon coupling strength. However this is strongly dependent on the magnetic fluctuation model assumed and also on the 
properties of the Coma cluster. Assuming the current available data on the Coma cluster properties \cite{Feretti95} we find,
$$g_{\rm eff}^{\rm (cell)} < 2.2\times 10^{-8}\,{\rm GeV}^{-1}$$
using the cell model, and
$$
g_{\rm eff}^{\rm (Kolmo)} < (7.2 \text{--} 32.5) \times 10^{-10}\,{\rm GeV}^{-1},
$$
assuming a Kolmogorov turbulence model. We note that other clusters such as Hydra A, a good example of a cooling-core cluster, have approximately similar conversion probabilities,
since although the observed magnetic field strength at the centre of the cluster is much greater (enhances chameleon effect) the electron density is also much higher (supresses chameleon effect).

The model dependence of the effective coupling strength leads us to 
reevaluate the constraints given in Ref. \cite{Burrage08}. Assuming a cell model 
for the magnetic field fluctuations the constraint on the coupling 
strength from starlight polarization was found to be $g_{\rm eff} < 9.1 
\times 10^{-10} {\rm GeV}^{-1}$. From our analysis we have found that the 
conversion probability increases by a factor of 
$\left(\bar{\Delta}/N\right)^{1/3}$ if a Kolmogorov turbulence model is 
assumed instead. This increases the sensitivity to $g_{\rm eff}$ by 
approximately $\left(\bar{\Delta}/N\right)^{1/6}$. In our galaxy, 
$\bar{\Delta}/N \sim 50\text{--}100$, for the objects considered in Ref. 
\cite{Burrage08}. Thus we expect to increase the constraint to approximately 
$g_{\rm eff} < 4\times 10^{-10} {\rm GeV}^{-1}$ if a Kolmogorov turbulence model is used 
in the analysis of the starlight polarization data. We note that the 
current constraints coming from starlight polarization are tighter than 
those from SZ measurements. However for the starlight polarization the data 
is predominantly limited by intrinsic scatter and improvements are likely 
to be statistical, while there is the possibility of SZ observations away 
from the cluster core leading to even better SZ constraints (see 
below). It should be mentioned, however, that there is significant 
uncertainty in the SZ constraints coming from a lack of detailed knowledge 
about cluster magnetic fields on scales $\sim k_{\rm crit}^{-1}$, whereas the 
starlight polarization analysis rests on knowledge of the power spectrum of 
the galaxy magnetic field and electron density fluctuations which are much 
better known.

In addition to the frequency dependence of the chameleon effect, it is possible to exploit the dependence on magnetic field strength and electron density to examine the radial
dependence within the galaxy cluster. The magnetic field and electron density are expected to decrease with radius. We have assumed a simple scaling for the
electron density following a $\beta$-profile, $n_{e} = n_{0}\left(1+\frac{r^{2}}{r_{\rm c}^{2}}\right)^{-3\beta/2}$, with $B \propto n_{e}^{\eta}$. In Fig.
\ref{fig:profile} we plotted both the thermal SZ radial profile and the predicted chameleonic radial profile for a hypothetical cluster with chameleon-photon conversion
probability  $\Pphi(204\,{\rm GHz}) = 3\times 10^{-5}$ through the centre of the cluster. We found that the chameleonic effect generally dominates towards the edges of the cluster.
The suggestion of a discrepancy between the inferred SZ effect from X-ray emission data and the WMAP SZ measurements \cite{ROSATvsWMAP} could potentially be attributed to a 
chameleon, and this requires further investigation. In any case, future measurements of the SZ profile at higher frequencies would lead to even better constraints on the chameleon model.

Finally we looked at the modification to the CMB polarization. We found that the contribution to the modification from chameleon-photon mixing arising from the random component of
 the intra-cluster magnetic field, i.e. that tangled on scales much smaller than the size of the cluster, is severly limited by the angular and spectral resolution of the 
 detector. We would require angular resolution on the scale of the magnetic
domains in the cluster, approximately one-hundredth the size of the
cluster, and frequency resolution of the order $\delta \lambda / \lambda \leq 10^{-6}$ before any effect could be detected in the CMB polarization. The contribution arising from 
the regular component of the magnetic field is more significant and potentially detectable. So far polarization measurements of the CMB in the direction of galaxy clusters have 
not been made. The upcoming SPT-pol experiment \cite{SPTelescope} will change this. Assuming a Kolomogorov turbulence model, we predict the change to the fractional CMB linear 
polarization:
\be 
\Delta Q \approx \Delta U &\approx& (0\text{--}1.5 \times 10^{-9}) g_{10}^2 \nu_{214}^{5/3}B_{1 \mu{\rm G}}^2 L_{200} \nonumber \\
&& n_{0.01}^{-5/3} \left(L_{1\kpc}^{\rm coh}\right)^{-2/3}, \nonumber
\ee
which for some clusters is approaching the level of the intrinsic CMB polarisation.

We conclude by mentioning two points which at first may appear to be potential sources of error in our predictions. Firstly we remind the reader that 
the magnetic field strengths of most galaxy clusters are determined
from Faraday rotation measures, looking at the degree to which the
polarization plane of some background source is rotated due to the
presence of the cluster magnetic field. It is feasible that the
presence of a chameleon could influence the change to the polarisation angle and hence the inferred magnetic field strength of these clusters. However
the Faraday RMs generally come from low frequency measurements around $10\,\mathrm{GHz}$ for which the chameleon interaction is greatly
suppressed. Therefore we believe it is realistic to ignore the
contamination when considering the cluster magnetic field strength. 

Secondly, in these predictions we have neglected the contribution from
our own galactic magnetic field. In \cite{Burrage08} the bound on the chameleon-photon conversion probability in our own galaxy coming from starlight polarization was found to be
$\Pbar\lesssim \text{few}\,\times 10^{-6}$ at $\omega = 1\,{\rm eV}$. Assuming the probability scales as $\omega^{5/3}$, then the contribution at CMB frequencies from the galactic
magnetic field would be less than
$\mathcal{O}(10^{-10})$, which is negligble.   

\vspace{0.5cm}

\noindent{\bf Acknowledgements:} ACD, CAOS and DJS are supported by
STFC. We are grateful to Clare Burrage, Anthony Challinor, Keith Grainge, Hiranya Peiris and Paul Shellard for extremely helpful conversations. 

\appendix
\section{Photon-Scalar Mixing}
In the weak mixing limit, we saw in \S \ref{sec:Optics} that the photon-scalar mixing can be described by a function $A_{i}(z)$ where $i = \left\lbrace x,y\right\rbrace$, and
\be
A_{i}(z) = \int_{0}^{z}\dd x \frac{B_{i}(x)}{2M_{\rm eff}}e^{2i\Delta(x)}.\label{AiEqn}
\ee
We assume that $\mathbf{B}$ has both a regular component, $\mathbf{B}_{\rm reg}$, which we take to be constant,  and a random component, $\delta \mathbf{B}$.  We also found in 
\S \ref{sec:Optics} that the probability of a photon converting into a scalar field, $\Pphi$, and the induced contributions, $\mathcal{Q}_{\rm q}$, $\mathcal{Q}_{\rm u}$ and 
$\mathcal{Q}_{\rm v}$ to the Stokes parameters $Q$, $U$ and $V$, are given in terms of the $A_{i}$ thus:
\be
\Pphi(z) &=& \frac{1}{2}\left(\vert A_{x}(z)\vert^2 + \vert A_{y}(z)\vert^2\right), \nonumber \\
\mathcal{Q}_{\rm q}(z) &=& \frac{1}{2}\left(\vert A_{x}(z)\vert^2 - \vert A_{y}(z)\vert^2\right), \nonumber \\
\mathcal{Q}_{\rm u}(z) &=& {\rm Re}(A_{x}^{\ast}(z)A_{y}(z)), \nonumber \\
\mathcal{Q}_{\rm v}(z) &=& {\rm Im}(A_{x}^{\ast}(z)A_{y}(z)).\nonumber
\ee
We now evaluate these quantities in the cell and power spectrum models for the magnetic field fluctuations when $z=L$, where $L$ is the path length through the magnetic region.

\subsection{The Cell Model}\label{app:Cell}
 In the cell model for the magnetic field we assume that along the
 light path, $0< z < L$, the components of the random magnetic field $\delta \mathbf{B}$ perpendicular to the line of sight are given by
\be
(\delta \mathbf{B})_{x} = B_{\rm rand}\cos \theta_{n}, \nonumber\\
(\delta \mathbf{B})_{y} = B_{\rm rand}\sin \theta_{n},\nonumber
\ee
in the region $(n-1)L_{\rm coh} < z < n L_{\rm coh}$, where $L_{\rm
 coh}$ is the coherence length and $\theta_{n} \sim U(0,2\pi]$. We
 define $N=L/L_{\rm coh}$ with $n$ running from 1 to $N$.  We take
 $B_{\rm rand}$ to be fixed and assume a constant value of $n_{\rm e}=
 \bar{n}_{\rm e}$. Thus $\Delta(x) = \bar{\Delta} x /L$, where 
\be
\bar{\Delta} \equiv \frac{\bar{m}_{\rm eff}^2L}{4\omega}, \nonumber
\ee
and $\bar{m}_{\rm eff}^2 \simeq -\omega_{\rm pl}^2= - (4\pi
 \alpha_{\rm em} \bar{n}_{e}/m_{\rm e})^2$ in the cluster atmosphere. It is now straightforward to evaluate $A_{i}$ and we find,
\be
A_{i}(L) &=&  (\mathbf{A}_{\rm reg})_{i} + (\delta \mathbf{A})_{i}, \nonumber\\
\mathbf{A}_{\rm reg} &=& \frac{2\mathbf{B}_{\rm reg} \omega e^{i\bar{\Delta}}}{M_{\rm eff}m_{\rm eff}^2}\sin\bar{\Delta}, \nonumber\\
\delta \mathbf{A} &=& \frac{2B_{\rm rand} \omega e^{\frac{i\bar{\Delta}}{N}}}{M_{\rm eff}m_{\rm eff}^2}\sin\left(\frac{\bar{\Delta}}{N}\right) \sum_{r = 0}^{N-1} 
\hatb{n}^{(r)} e^{\frac{2i \bar{\Delta} r}{N}}, \nonumber
\ee
where $\hatb{n}^{(n)} = (\cos \theta_n,\sin \theta_n,0)^{\rm T}$.   Taking $(\mathbf{B}_{\rm reg})_{x} = B_{\rm reg}\cos \theta_{\rm reg}$ and  $(\mathbf{B}_{\rm reg})_{y} = 
B_{\rm reg}\sin \theta_{\rm reg}$, we find
\be
\Pphi(L) = \mathcal{P}^{\rm reg} + \mathcal{P}^{\rm cross} + \mathcal{P}^{\rm rand}, \nonumber\\
\mathcal{Q}_{i}(L) = \mathcal{Q}^{\rm reg}_{i} + \mathcal{Q}^{\rm cross}_{i} + \mathcal{Q}^{\rm rand}_{i}, \nonumber
\ee
where the regular components are given by:
\be
\mathcal{P}^{\rm reg} &=& \frac{1}{2}\left(\frac{2 B_{\rm reg} \omega }{M_{\rm eff}m_{\rm eff}^2}\right)^2\sin^2\bar{\Delta}, \nonumber\\
\mathcal{Q}^{\rm reg}_{\rm q} &=&  \frac{\cos 2\theta_{\rm reg}}{2}\left(\frac{2 B_{\rm reg} \omega }{M_{\rm eff}m_{\rm eff}^2}\right)^2\sin^2\bar{\Delta}, \nonumber\\
\mathcal{Q}^{\rm reg}_{\rm u} &=& \frac{\sin 2\theta_{\rm reg}}{2}\left(\frac{2 B_{\rm reg} \omega }{M_{\rm eff}m_{\rm eff}^2}\right)^2\sin^2\bar{\Delta}, \nonumber\\
\mathcal{Q}^{\rm reg}_{\rm v} &=& 0.\nonumber
\ee
Similarly the cross-terms are given by:
\be
\mathcal{P}^{\rm cross} &=& \left(\frac{2 \omega }{M_{\rm eff}m_{\rm eff}^2 }\right)^2B_{\rm reg} B_{\rm rand}\sin \bar{\Delta}\sin \left(\frac{\bar{\Delta}}{N}\right)\nonumber \\ &&\sum_{r=0}^{N-1} \cos(\theta_r-\theta_{\rm reg}) \cos\left(\frac{\bar{\Delta}}{N}(2r+1-N)\right), \nonumber\\
\mathcal{Q}^{\rm cross}_{\rm q} &=&  \left(\frac{2 \omega }{M_{\rm eff}m_{\rm eff}^2 }\right)^2B_{\rm reg} B_{\rm rand}\sin \bar{\Delta}\sin \left(\frac{\bar{\Delta}}{N}\right)\nonumber\\ &&\sum_{r=0}^{N-1} \cos(\theta_r+\theta_{\rm reg}) \cos\left(\frac{\bar{\Delta}}{N}(2r+1-N)\right), \nonumber\\
\mathcal{Q}^{\rm cross}_{\rm u} &=&  \left(\frac{2 \omega }{M_{\rm eff}m_{\rm eff}^2 }\right)^2B_{\rm reg} B_{\rm rand}\sin \bar{\Delta}\sin \left(\frac{\bar{\Delta}}{N}\right)\nonumber \\ &&\sum_{r=0}^{N-1} \sin(\theta_r+\theta_{\rm reg}) \cos\left(\frac{\bar{\Delta}}{N}(2r+1-N)\right), \nonumber \\
\mathcal{Q}^{\rm cross}_{\rm v} &=& \left(\frac{2 \omega }{M_{\rm eff}m_{\rm eff}^2 }\right)^2B_{\rm reg} B_{\rm rand}\sin \bar{\Delta}\sin \left(\frac{\bar{\Delta}}{N}\right)\nonumber \\ &&\sum_{r=0}^{N-1} \sin(\theta_r-\theta_{\rm reg}) \sin\left(\frac{\bar{\Delta}}{N}(2r+1-N)\right).\nonumber
\ee
Finally the terms due entirely to the random component of the magnetic field are:
\be
\mathcal{P}^{\rm rand} &=& \frac{1}{2}\left(\frac{2 B_{\rm rand} \omega }{M_{\rm eff}m_{\rm eff}^2}\right)^2\sin^2 \left(\frac{\bar{\Delta}}{N}\right)\nonumber\\ &&\sum_{s=0}^{N-1} \sum_{r=0}^{N-1} \cos(\theta_{r}-\theta_s) \cos\left(\frac{2\bar{\Delta}}{N}(r-s)\right), \nonumber \\
\mathcal{Q}^{\rm rand}_{\rm q} &=&  \frac{1}{2}\left(\frac{2 B_{\rm rand} \omega }{M_{\rm eff}m_{\rm eff}^2}\right)^2\sin^2 \left(\frac{\bar{\Delta}}{N}\right)\nonumber\\ &&\sum_{s=0}^{N-1} \sum_{r=0}^{N-1} \cos(\theta_{r}+\theta_s) \cos\left(\frac{2\bar{\Delta}}{N}(r-s)\right), \nonumber\\
\mathcal{Q}^{\rm rand}_{\rm u} &=& \frac{1}{2}\left(\frac{2 B_{\rm rand} \omega }{M_{\rm eff}m_{\rm eff}^2}\right)^2\sin^2 \left(\frac{\bar{\Delta}}{N}\right)\nonumber\\ &&\sum_{s=0}^{N-1} \sum_{r=0}^{N-1} \sin(\theta_{r}+\theta_s) \cos\left(\frac{2\bar{\Delta}}{N}(r-s)\right), \nonumber \\
\mathcal{Q}^{\rm rand}_{\rm v} &=& \frac{1}{2}\left(\frac{2 B_{\rm rand} \omega }{M_{\rm eff}m_{\rm eff}^2}\right)^2\sin^2 \left(\frac{\bar{\Delta}}{N}\right)\nonumber\\ &&\sum_{s=0}^{N-1} \sum_{r=0}^{N-1} \sin(\theta_{r}-\theta_s) \sin\left(\frac{2\bar{\Delta}}{N}(r-s)\right).\nonumber
\ee
Taking an average over many lines of sight through the cluster is equivalent to averaging over the $\theta_{n}$.  Doing this it is straightforward to check that the average values
of $\Pphi$ and the $\mathcal{Q}_{i}$ are given by,
\be
\Pbar \equiv \left \langle \Pphi \right\rangle &=& \frac{1}{2}\left(\frac{2 B_{\rm reg} \omega }{M_{\rm eff}m_{\rm eff}^2}\right)^2\sin^2\bar{\Delta} \nonumber \\&&+ \frac{N}{2}\left(\frac{2B_{\rm rand} \omega}{M_{\rm eff}m_{\rm eff}^2}\right)^2\sin^2\left(\frac{\bar{\Delta}}{N}\right), \nonumber  \\
\bar{\mathcal{Q}}_{i} \equiv \left \langle \mathcal{Q}_{i} \right\rangle &=& \mathcal{Q}^{\rm reg}_{i}. \nonumber
\ee
We can also calculate the variance of the $\mathcal{Q}_{i}$ over many
 lines of sight. Assuming $\vert\bar{\Delta}\vert/N \gg 1$, $N \gg 1$
 and remembering $B_{\rm rand}\sim\mathcal{O}(B_{\rm reg})$, we find
\be
\sigma_{i}^2 & =& \left\langle \mathcal{Q}_{i}^2\right\rangle -
 \bar{\mathcal{Q}}_{i}^2 \approx \frac{1}{2}\Pbar^2\,. \nonumber
\ee
This assumes that measurements of the Stokes parameters, and
 hence the $\mathcal{Q}_{i}$, are exactly monochromatic.   The
 observable $\sigma_{i}^2$ can be greatly suppressed relative to the
 above estimate when, as is always the case in practice, measurements
 actually comprise an average of the observable quantity over a
 frequency band about the desired measurement frequency.  We discuss
 this further in \S \ref{app:Measurement} below.

\subsection{The Power Spectrum Model}\label{app:Powerspectrum}
In the power spectrum model, we assume that fluctuations in
$\mathbf{B}$ occur on all spatial scales, and we describe the magnitude
of the fluctuations on different scales by a power spectrum, $P_{\rm
  B}(k)$.  We also allow fluctuations in the electron number density,
$n_{\rm e}$, which are described by another power spectrum, $P_{\rm
  N}(k)$.  We define $n_{\rm e}(z) = \bar{n}_{\rm e}(1+\delta_{\rm
  n}(z))$, where $\bar{n}_{e}$ is the average electron density along
the path through the magnetic region.  We also define,
$$
\bar{\delta}_{\rm n}(z)=\frac{1}{z}\int_{0}^{z} \delta_{\rm n}(x) \dd x\,.
$$
Note $\bar{\delta}_{\rm n}(L)=0$.  With $A_{i}$ defined by
Eq. (\ref{AiEqn}), we define $Z = (1+\bar{\delta}_{\rm n}(z))z$ and
$\bar{\Delta} = \bar{m}_{\rm eff}^2L/4\omega\propto -\bar{n}_{\rm e}L /4\omega$. It
follows that $\Delta(z) = \bar{\Delta}Z/L$.  We assume that
$\vert\bar{\delta}_{\rm n}(z)\vert \ll 1$ along the majority of the path, and
so $B_{i}(z) \approx B_{i}(Z)$. We then have,
\be
A_{i}(\bar{\Delta}) \approx \int_{0}^{L}  \dd Z\,\frac{B_{i}(Z\hatb{z})}{2M_{\rm eff}(1+\delta_{\rm n}(Z\hatb{z}))}e^{\frac{2i\bar{\Delta}Z}{L}}. \nonumber
\ee
To calculate the expected values of $\Pphi$ and the
$\mathcal{Q}_{i}$, we must calculate the correlation at different frequencies:
\be
&\mathcal{A}_{ij}(\bar{\Delta}_{1},\bar{\Delta}_{2}) \equiv \left\langle A_{i}(\bar{\Delta}_{1})A_{j}^{\ast}(\bar{\Delta}_{2})\right\rangle &\nonumber\\
&= \frac{1}{4M_{\rm eff}^2}\int_{0}^{L}\dd x\int_{0}^{L}\dd y R_{ij}^{\rm tot}((x-y)\hatb{z})e^{2i(\bar{\Delta}_{1}x-\bar{\Delta}_{2}y)/L}&, \nonumber
\ee
where the correlation function is defined,
$$
R_{ij}^{\rm tot}(x\hatb{z}) \equiv \left\langle \frac{B_{i}(y\hatb{z})B_{j}((y+x)\hatb{z})}{(1+\delta_{\rm n}(y\hatb{z})(1+\delta_{\rm n}((y+x)\hatb{z})}\right\rangle.
$$
We split the magnetic field into a regular and turbulent component:
$\mathbf{B}=\mathbf{B}_{\rm reg}+\delta\mathbf{B}$. Assuming isotropy of the fluctuations in $\mathbf{B}$ and $n_{\rm e}$, we rewrite $R_{ij}^{\rm tot}$ thus,
\be
R_{ij}^{\rm tot}(x\hatb{z}) &=& B^{\rm reg}_{i}B^{\rm reg}_{j} \nonumber \\ &&\cdot\left\langle (1+\delta_{\rm n}(y\hatb{z}))^{-1}(1+\delta_{\rm n}((y+x)\hatb{z}))^{-1}\right\rangle \nonumber \\
&&+\frac{\delta_{ij}}{3}\left\langle \frac{\delta \mathbf{B}(y\hatb{z}) \cdot \delta \mathbf{B}((y+x)\hatb{z})}{(1+\delta_{\rm n}(y\hatb{z})(1+\delta_{\rm n}((y+x)\hatb{z}))}\right\rangle. \nonumber
\ee
We note that isotropy of the magnetic fluctuations implies that the off-diagonal components of
$R_{ij}^{\rm tot}$ are sourced only by the regular magnetic field.  We define a combined power spectrum $P^{\rm tot}_{ij}(k)$ by,
\be
R_{ij}^{\rm tot}(x\hatb{z}) &=& \frac{1}{4\pi}\int \dd^3 k\,
P_{ij}^{\rm tot}(k)e^{2\pi ix\mathbf{k}\cdot \hatb{z}}. \nonumber
\ee
We then have,
\be
\mathcal{A}_{ij} = \frac{L^2}{4M_{\rm eff}^2}e^{i(\bar{\Delta}_{1}-\bar{\Delta}_{2})}\int_{0}^{\infty} k^2 P_{ij}^{\rm tot}(k)K(\pi kL;\bar{\Delta}_1,\bar{\Delta}_2)\dd k,\nonumber
\ee
where, 
\be
K(\pi kL;\bar{\Delta}_{1},\bar{\Delta}_{2}) = \int_{0}^{1} \dd s & &
\frac{\sin(2\pi kL s)}{2\pi kL s}
\frac{\sin(2\bar{\Delta}_{-}(1-s))}{\bar{\Delta}_{-}} \nonumber \\
& & \cdot \cos (2\bar{\Delta}_{+}s), \nonumber
\ee
and we have defined $ 2\bar{\Delta}_{\pm} \equiv \bar{\Delta}_{2} \pm\bar{\Delta}_{1}$. This integral may be evaluated explicitly in terms of the functions:
\be
\Cin(x) &=& \int_{0}^{x} \frac{\sin^2 (y/2)}{y/2} \dd y, \nonumber \\
\Si(x) &=& \int_{0}^{x} \frac{\sin y}{y} \dd y.\nonumber
\ee
We define $\delta \Cin(x,y) = \Cin(x+y)-\Cin(x-y)$ and $\delta \Si(x,y) = \Si(x+y)-\Si(x-y)$.   We then have,
\be
4xK(x;\bar{\Delta}_{1},\bar{\Delta}_{2}) &=& \frac{\cos 2\bar{\Delta}_{-}}{2\bar{\Delta}_{-}}\left[\delta\Cin(2\bar{\Delta}_{1},2x) \right. \nonumber\\ &&\left. -\delta\Cin(2\bar{\Delta}_{2},2x)\right] \nonumber\\
&& + \frac{\sin 2\bar{\Delta}_{-}}{2\bar{\Delta}_{-}}\left[\delta\Si(2\bar{\Delta}_{1},2x)\right. \nonumber \\ && \left.+\delta\Si(2\bar{\Delta}_{2},2x)\right]. \nonumber
\ee
We are concerned with CMB radiation, for which $\omega \sim
10^{-5}\text{--}10^{-3}\eV$, propagating over distances of the order
of $100\kpc$ through clusters.  It is easy to see that for such a scenario $\vert\bar{\Delta}\vert \gg 1$.  We evaluate $K(x;\bar{\Delta}_{1},\bar{\Delta}_{2})$ in the asymptotic
 limit where $\vert \bar{\Delta}_{1}\vert, \vert \bar{\Delta}_{2}\vert \gg 1$. When $\min(\vert \bar{\Delta}_1\vert, \vert\bar{\Delta}_2\vert) \gg 1$ and 
 $x \ll \min(\vert \bar{\Delta}_1\vert, \vert\bar{\Delta}_2\vert)$ we have,
\be
4xK(x;\bar{\Delta}_1,\bar{\Delta}_2) &\simeq & \frac{2x \cos
  2\bar{\Delta}_{-}}{\bar{\Delta}_1 \bar{\Delta}_2} \nonumber\\ && - \frac{\sin 2x \cos 2\bar{\Delta}_{+}}{\bar{\Delta}_1\bar{\Delta}_2},\nonumber
\ee
and when $x \gg \min(\vert \bar{\Delta}_1\vert, \vert\bar{\Delta}_2\vert)$, we have
\be
4xK(x;\bar{\Delta}_1,\bar{\Delta}_2) &\simeq& \frac{2\pi \sin 2\bar{\Delta}_{-}}{2\bar{\Delta}_{-}}-\frac{2\cos 2\bar{\Delta}_{-}}{x}.\nonumber
\ee 
We define $k_{i} = \vert \bar{\Delta}_i \vert/\pi L$. We assume that
the fluctuations in $\mathbf{B}$ and $n_{\rm e}$ are such that $P^{\rm tot}_{ij}(k)$ drops off faster than $k^{-3}$ for all $k >k_{\ast}$ where $k_{\ast} \ll \min(k_{i})$. 
This is equivalent to assuming that the dominant contribution to $R^{\rm tot}_{ij}(0)$ comes from spatial scales than are much larger than $k_{i}^{-1}$.  We expect that this 
dominant contribution will come from scales of the order of the coherence lengths of the magnetic field and electron density fluctuations ($L_{\rm B}$ and $L_{\rm N}$ 
respectively) and so we are assuming that $k_{i}^{-1} \ll L_{\rm B},\,L_{\rm N}$.   With this assumption, in the limit of large $\vert \bar{\Delta}_{i}\vert$, we have to leading 
order,
\be
e^{2i\bar{\Delta}_{-}}4M_{\rm eff}^2\mathcal{A}_{ij} &\simeq& \frac{L^2\cos 2\bar{\Delta}_{-}}{2\bar{\Delta}_1\bar{\Delta}_2} \int_{0}^{\infty} k^2 P_{ij}^{\rm tot}(k)\dd k \nonumber \\
&& - \frac{L \cos 2\bar{\Delta}_{+}}{4\pi \bar{\Delta}_1\bar{\Delta}_2} \int_{0}^{\infty} k \sin(2\pi kL) P_{ij}^{\rm tot}(k)\dd k \nonumber \\
&& + \frac{L\sin 2\bar{\Delta}_{-}}{8\bar{\Delta}_{-}} \left[\int_{k_{1}}^{\infty} k P_{ij}^{\rm tot}(k)\dd k  \right. \nonumber \\
&& + \left. \int_{k_{2}}^{\infty} k P_{ij}^{\rm tot}(k)\dd k \right].\nonumber
\ee
We define,
$$
W_{ij}(k) = \int_{k}^{\infty} qP_{ij}^{\rm tot}(q) \dd q, 
$$
and we can then rewrite $\mathcal{A}_{ij}$ as,
\be
e^{2i\bar{\Delta}_{-}}\mathcal{A}_{ij} &\simeq& \frac{4 \omega_{1}\omega_{2}}{2M_{\rm eff}^2 \bar{m}_{\rm eff}^4}R_{ij}^{\rm tot}(0)\cos 2\bar{\Delta}_{-} \nonumber\\
&&- \frac{4 \omega_{1}\omega_{2}}{2M_{\rm eff}^2 \bar{m}_{\rm eff}^4} R_{ij}^{\rm tot}(L\hatb{z}) \cos 2\bar{\Delta}_{+} \nonumber \\
&&+\frac{L\sin 2\bar{\Delta}_{-}}{32M_{\rm eff}^2\bar{\Delta}_{-}} \left[W_{ij}(k_1)+W_{ij}(k_2)\right]. \nonumber
\ee
For simplicity, we assume that fluctuations in the magnetic field and electron density are uncorrelated. Thus
\be
&\left\langle  \frac{\delta\mathbf{B}(\mathbf{y})\cdot  \delta \mathbf{B}(\mathbf{y}+\mathbf{x})}{(1+\delta_{\rm n}(\mathbf{y}))(1+\delta_{\rm n}(\mathbf{x}+\mathbf{y}))}\right\rangle 
= \left\langle \delta\mathbf{B}(\mathbf{y})\cdot  \delta \mathbf{B}(\mathbf{y}+\mathbf{x})\right\rangle & \nonumber\\ 
&\cdot\left\langle (1+\delta_{\rm n}(\mathbf{y}))^{-1}(1+\delta_{\rm n}(\mathbf{y}+\mathbf{x}))^{-1}\right\rangle. &\nonumber
\ee
This implies that,
\be
R_{ij}^{\rm tot}(0) \approx \left(B_{i}^{\rm reg}B_{j}^{\rm reg} + \frac{1}{3}\delta_{ij}\left\langle \delta \mathbf{B}^2\right\rangle\right)\left\langle
\frac{\bar{n}_{\rm e}^2}{n_{\rm e}^2}\right\rangle. \nonumber
\ee
We assume that fluctuations in $n_{\rm e}$ follow an approximately log-normal distribution, and hence
\be
\left\langle\frac{\bar{n}_{\rm e}^2}{n_{\rm e}^2}\right\rangle \approx\left\langle\frac{n_{\rm e}^2}{\bar{n}_{\rm e}^2}\right\rangle^3 \equiv I_{\rm N}^3\,. \nonumber
\ee
We assume, as is the case in the situations we consider, that $L$ is much larger than the correlation length of the magnetic and electron density fluctuations.  
This implies that the only contribution to $R^{\rm tot}_{ij}(L\hatb{z})$ which may result in a leading order contribution to $\mathcal{A}_{ij}$ is,
\be
R_{ij}^{\rm tot}(L\hatb{z}) \sim B_{i}^{\rm reg}B_{j}^{\rm reg}. \nonumber
\ee
Finally, we must evaluate $W_{ij}(k)$ in terms of the magnetic and electron density power spectra.  We define,
\be
R_{\delta}(\mathbf{x})&=&\left\langle (1+\delta_{\rm n}(\mathbf{y}))^{-1}(1+\delta_{\rm n}(\mathbf{y}+\mathbf{x}))^{-1}\right\rangle, \nonumber\\
R_{\rm B}(\mathbf{x}) &=& \left\langle \delta\mathbf{B}(\mathbf{y}) \cdot \delta \mathbf{B}(\mathbf{y}+\mathbf{x})\right\rangle.\nonumber
\ee
We define $P_{\rm B}(k)$ and $P_{\rm \delta}(k)$ relative to $R_{\rm B}$ and $R_{\delta}$ by the relation,
\be
R(\mathbf{x}) = \frac{1}{4\pi} \int \dd^3 k \,P(k) e^{2\pi i\mathbf{k}\cdot \mathbf{x}}.\nonumber
\ee
We also define,
\be
R_{\rm N}(\mathbf{x})&=&\left\langle \delta n_{\rm e}(\mathbf{y})\delta n_{\rm e}(\mathbf{y}+\mathbf{x})\right\rangle, \nonumber
\ee
and corresponding Fourier transform, $P_{\rm N}$. We separate the fluctuations in $n_{\rm e}$  into short and long wavelength fluctuations, $\delta_s$ and $\delta_l$ respectively, and 
assume they are approximately independent. Thus $n_{\rm e}=\bar{n}_{\rm e}(1+\delta_l)(1+\delta_s)$.  This parameterisation is consistent with the assumption that $n_{\rm e}/\bar{n}_{\rm e}$ has an 
 approximately log-normal distribution. We assume that the short wavelength fluctuations are linear up to some cut-off scale 
 $k_{\rm lin}^{-1}$. Above this spatial scale we have the long wavelength fluctuations, which are not necessarily linear. We then have that over small spatial scales, $\ll k^{-1}_{\rm lin}$,
\be
R_{\delta}(\mathbf{x}) &\approx& \left\langle \frac{\bar{n}_{\rm e}^2}{n_{\rm e}^2}\right\rangle \left\langle (1+\delta_{s}(\mathbf{y}))^{-1}(1+\delta_{s}(\mathbf{y}+\mathbf{x}))^{-1}\right\rangle,\nonumber  \\
&\approx& \left\langle \frac{\bar{n}_{\rm e}^2}{n_{\rm e}^2}\right\rangle\left\langle \frac{n_{\rm e}^2}{\bar{n}_{\rm e}^2}\right\rangle^{-1} \left[1+ n_{\rm e}^{-2} R_{\rm N}(\mathbf{x})\right], \nonumber \\
&=& I_{\rm N}^2 \left[ 1 + \bar{n}_{\rm e}^{-2}R_{\rm N}(\mathbf{x})\right]. \nonumber
\ee
It follows that for $k \gg k_{\rm lin}$ we have approximately,
\be
P_{\delta}(k) \approx I_{\rm N}^2 \bar{n}_{\rm e}^{-2} P_{\rm N}(k). \nonumber
\ee
We assume that $k_{\rm lin} \ll k_{1}, \, k_{2}$.  Finally we define,
\be
P_{\rm B\delta}(k) = P_{\rm B\delta}(\mathbf{k}) &=& \frac{1}{4\pi} \int \dd^3 p P_{\delta}(p)P_{\rm B}(\Vert \mathbf{k}-\mathbf{p}\Vert).\nonumber 
\ee
We note that,
\be
&\frac{1}{4\pi}\int \dd^3k P_{\rm B\delta}(k)e^{2\pi i\mathbf{k}\cdot\mathbf{x}} = \left\langle \delta\mathbf{B}(\mathbf{y})
\cdot  \delta \mathbf{B}(\mathbf{y}+\mathbf{x})\right\rangle & \nonumber\\ &\cdot\left\langle (1+\delta_{n}(\mathbf{y}))^{-1}(1+\delta_{n}(\mathbf{y}+\mathbf{x}))^{-1}\right\rangle. &\nonumber
\ee
From the definition of $P_{ij}^{\rm tot}$ and $W_{ij}(k)$ we have,
\be
W_{ij}(k) &=& B_{i}^{\rm reg} B_{j}^{\rm reg}\int_{k}^{\infty}q P_{\delta}(q)\dd q \nonumber\\
&&+\frac{1}{3}\delta_{ij}\int_{k}^{\infty}q P_{B\delta}(q)\dd q. \nonumber
\ee
All that remains is to evaluate,
\be
\int_{k}^{\infty} q P_{\rm B\delta}(q) \dd q = \int \dd^3 p \frac{P_{\delta}(p)}{4\pi} \int_{V_{k}(\mathbf{p})} \dd^3 r \frac{P_{\rm B}(r)}{4\pi \Vert \mathbf{r}+\mathbf{p}\Vert}, \nonumber
\ee
where $V_{k}(\mathbf{p})$ is defined by $\Vert \mathbf{r}+\mathbf{p}\Vert > k$.  We approximate this integral using $\Vert \mathbf{r} + \mathbf{p}\Vert 
\approx p$ in $0 < r < p$ and $\Vert \mathbf{r} + \mathbf{p}\Vert \approx r$ in $0 < p < r$, and find
\be
\int_{k}^{\infty} q P_{B\delta}(q)\dd q &\approx& \left\langle \delta \mathbf{B}^2 \right\rangle \int_{k}^{\infty}q P_{\delta}(q)\dd q \nonumber \\ &&+ I_{N}^3 \int_{k}^{\infty}q
 P_{B}(q)\dd q. \nonumber
\ee
Defining,
\be
W_{\rm B/N}(k) = \int_{k}^{\infty} q\dd q P_{\rm B/N}(q), \nonumber
\ee
we have for $\mathcal{A}_{ij}$,
\be
e^{2i\bar{\Delta}_{-}}\mathcal{A}_{ij}(\bar{\Delta}_1,\bar{\Delta}_2) &\simeq& \frac{4 I_{N}^3 B_{ij}^2 \omega_{1}\omega_{2}}{2M_{\rm eff}^2 \bar{m}_{\rm eff}^4}\cos
2\bar{\Delta}_{-} \nonumber\\
&&- \frac{4 B_{i}^{\rm reg} B_{j}^{\rm reg} \omega_{1}\omega_{2}}{2M_{\rm eff}^2 \bar{m}_{\rm eff}^4}\cos 2\bar{\Delta}_{+} \nonumber \\
&&+\frac{B_{ij}^2 I_{\rm N}^2 L {\rm sinc} 2\bar{\Delta}_{-}}{16M_{\rm eff}^2 \bar{n}_e^2} \left[W_{N}(k_1)+W_{N}(k_2)\right] \nonumber \\
&&+\frac{\delta_{ij} I_{\rm N}^3 L {\rm sinc} 2\bar{\Delta}_{-}}{48M_{\rm eff}^2} \left[W_{B}(k_1)+W_{B}(k_2)\right], \nonumber
\ee
where $B^{2}_{ij} = B_{i}^{\rm reg}B_{j}^{\rm reg} + \delta_{ij}\left\langle \delta \mathbf{B}^2\right\rangle /3$.  It is now straightforward to calculate the expectations of 
$\Pphi$ and the $\mathcal{Q}_{i}$ at a single frequency corresponding to $\bar{\Delta}$, since
\be
\Pbar &=& \frac{1}{2}\left(\mathcal{A}_{xx}(\bar{\Delta},\bar{\Delta})+\mathcal{A}_{yy}(\bar{\Delta},\bar{\Delta})\right),\nonumber  \\
\bar{\mathcal{Q}}_{\rm q} &=& \frac{1}{2}\left(\mathcal{A}_{xx}(\bar{\Delta},\bar{\Delta})-\mathcal{A}_{yy}(\bar{\Delta},\bar{\Delta})\right), \nonumber \\
\bar{\mathcal{Q}}_{\rm u} &=& \frac{1}{2}\left(\mathcal{A}_{xy}(\bar{\Delta},\bar{\Delta})+\mathcal{A}_{yx}(\bar{\Delta},\bar{\Delta})\right),\nonumber \\
\bar{\mathcal{Q}}_{\rm v} &=& \frac{i}{2}\left(\mathcal{A}_{xy}(\bar{\Delta},\bar{\Delta})-\mathcal{A}_{yx}(\bar{\Delta},\bar{\Delta})\right).\nonumber
\ee
Note that this implies $\bar{\mathcal{Q}}_{\rm v} = 0$. It follows that,
\be
\Pbar &\approx& \frac{1}{2}\left(\frac{2B_{\rm eff} \omega}{M_{\rm eff} \bar{m}_{\rm eff}^2}\right)^2 I_{\rm N}^3 \nonumber\\ 
&&- \frac{1}{4}\left(\frac{2B_{\rm reg} \omega}{M_{\rm eff} \bar{m}_{\rm eff}^2}\right)^2 \cos \left(2\bar{\Delta}\right) \nonumber 
\\ && + \frac{B_{\rm eff}^2 L }{8M_{\rm eff}^2 \bar{n}_{\rm e}^2} I_{\rm N}^2 W_{\rm N}(k_{\rm crit}) + \frac{L}{24M_{\rm eff}^2} I_{\rm N}^3  W_{\rm B}(k_{\rm crit}), \nonumber
\ee
where $B_{\rm eff}^2 \equiv B_{\rm reg}^2/2 + \left\langle \delta \mathbf{B}^2\right\rangle/3$, and $k_{\rm crit}=\vert \bar{\Delta} \vert/\pi L$. Similarly,
\be
\bar{\mathcal{Q}}_{\rm q} &\approx& \mathcal{Q}_{0}\cos 2\theta_{\rm reg}, \nonumber \\
\bar{\mathcal{Q}}_{\rm u} &\approx& \mathcal{Q}_{0}\sin 2\theta_{\rm reg}, \nonumber 
\ee
where,
\be
\mathcal{Q}_{0} &\equiv& \frac{1}{4}\left(\frac{2B_{\rm reg} \omega}{M_{\rm eff} \bar{m}_{\rm eff}^2}\right)^2 I_{\rm N}^3 \nonumber  \\ 
&&- \frac{1}{4}\left(\frac{2B_{\rm reg} \omega}{M_{\rm eff} \bar{m}_{\rm eff}^2}\right)^2 \cos \left(2\bar{\Delta}\right) \nonumber \\ 
&& + \frac{B_{\rm reg}^2 L }{16M_{\rm eff}^2 \bar{n}_{\rm e}^2} I_{\rm N}^2 W_{\rm N}(k_{\rm crit}). \nonumber
\ee
We can also calculate the variance of the $Q_{i}$.  We define,
\be
\sigma_{i}^2(\bar{\Delta}_1,\bar{\Delta}_2) = \left \langle
\mathcal{Q}_{i}(\bar{\Delta}_1)\mathcal{Q}_{i}(\bar{\Delta}_2)\right\rangle-\bar{\mathcal{Q}}_{i}(\bar{\Delta}_1)\bar{\mathcal{Q}}_{i}(\bar{\Delta}_2), \nonumber
\ee
and find that,
\be
4\sigma_{\rm q}^2 &=& \Vert \mathcal{A}_{xx}(\bar{\Delta}_1,\bar{\Delta}_2)\Vert^2 + \Vert \mathcal{A}_{xx}(\bar{\Delta}_1,-\bar{\Delta}_2)\Vert^2\nonumber\\
&+& \Vert \mathcal{A}_{yy}(\bar{\Delta}_1,\bar{\Delta}_2)\Vert^2+\Vert \mathcal{A}_{yy}(\bar{\Delta}_1,-\bar{\Delta}_2)\Vert^2  \nonumber \\
&-& \Vert \mathcal{A}_{xy}(\bar{\Delta}_1,\bar{\Delta}_2)\Vert^2 - \Vert \mathcal{A}_{xy}(\bar{\Delta}_1,-\bar{\Delta}_2)\Vert^2 \nonumber \\ 
&-&\Vert \mathcal{A}_{yx}(\bar{\Delta}_1,\bar{\Delta}_2)\Vert^2-\Vert \mathcal{A}_{yx}(\bar{\Delta}_1,-\bar{\Delta}_2)\Vert^2, \nonumber \\
4\sigma_{\rm u}^2 &=& 2{\rm Re}\left(\mathcal{A}_{xx}^{\ast}(\bar{\Delta}_1,\bar{\Delta}_2)\mathcal{A}_{yy}(\bar{\Delta}_1,\bar{\Delta}_2)\right) \nonumber\\
&+& 2{\rm Re}\left(\mathcal{A}_{xx}^{\ast}(\bar{\Delta}_1,-\bar{\Delta}_2)\mathcal{A}_{yy}(\bar{\Delta}_1,-\bar{\Delta}_2)\right) \nonumber \\
&+& 2{\rm Re}\left(\mathcal{A}_{xy}^{\ast}(\bar{\Delta}_1,\bar{\Delta}_2)\mathcal{A}_{yx}(\bar{\Delta}_1,\bar{\Delta}_2)\right) \nonumber \\
&+& 2{\rm Re}\left(\mathcal{A}_{xy}^{\ast}(\bar{\Delta}_1,-\bar{\Delta}_2)\mathcal{A}_{yx}(\bar{\Delta}_1,-\bar{\Delta}_2)\right), \nonumber \\ 
4\sigma_{\rm v}^2 &=& 2{\rm Re}\left(\mathcal{A}_{xx}^{\ast}(\bar{\Delta}_1,\bar{\Delta}_2)\mathcal{A}_{yy}(\bar{\Delta}_1,\bar{\Delta}_2)\right) \nonumber\\
&-& 2{\rm Re}\left(\mathcal{A}_{xx}^{\ast}(\bar{\Delta}_1,-\bar{\Delta}_2)\mathcal{A}_{yy}(\bar{\Delta}_1,-\bar{\Delta}_2)\right) \nonumber \\
&-& 2{\rm Re}\left(\mathcal{A}_{xy}^{\ast}(\bar{\Delta}_1,\bar{\Delta}_2)\mathcal{A}_{yx}(\bar{\Delta}_1,\bar{\Delta}_2)\right) \nonumber \\
&+& 2{\rm Re}\left(\mathcal{A}_{xy}^{\ast}(\bar{\Delta}_1,-\bar{\Delta}_2)\mathcal{A}_{yx}(\bar{\Delta}_1,-\bar{\Delta}_2)\right). \nonumber  
\ee
Note that we have assumed that the $A_{i}$ have an approximately Gaussian distribution so that the expectation of four $A_{i}$ can be written in terms of the expectation of two $A_{i}$.  
If we take $\bar{\Delta}_1 = \bar{\Delta}_2 = \bar{\Delta}$ then when $\Pbar$ is dominated by the terms proportional to $W_{\rm N}$, we have
\be
\sigma_{\rm q}^2 &\approx& \frac{1}{2}\Pbar^2+ \frac{1}{2}\mathcal{Q}_{0}^2 \cos 4\theta_{\rm reg}, \nonumber \\
\sigma_{\rm u}^2 &\approx& \frac{1}{2}\Pbar^2-\frac{1}{2}\mathcal{Q}_{0}^2 \cos 4\theta_{\rm reg},\nonumber \\
\sigma_{\rm v}^2 &\approx& \frac{1}{2}\Pbar^2-\frac{1}{2}\mathcal{Q}_{0}^2\,. \nonumber
\ee
We note that in the cell model we had $\sigma_{i}^2 \approx \Pbar/2$.

\subsection{Measurement Issues}\label{app:Measurement}
The above evaluation for the cell and power spectrum models is appropriate if measurements of the Stokes parameters and hence $\mathcal{Q}_{i}$ probe only a single frequency
 at a time, i.e. they are precisely monochromatic.  In practice, a measurement at a nominal frequency $\bar{\omega}$, will generally be an average over some frequency band with 
 width $\delta \omega$. We call $\delta \omega/\bar{\omega}$ the spectral resolution. We define $2\delta \Delta$ to be the difference in the value of $\Delta$ across the frequency
  band. When measurements of the Stokes parameters are made over a finite frequency band, the observable variance in the $\mathcal{Q}_{i}$ is no longer given by the 
  $\sigma_{i}^2$ expressions presented above but, in the (more general) case of the power spectrum model, by
\be
\hat{\sigma}^2_{i}(\bar{\Delta}) = \frac{1}{(2\delta \Delta)^2}\int_{-\delta \Delta}^{\delta \Delta} \dd \Delta_{1} \int_{-\delta \Delta}^{\delta \Delta} \dd \Delta_{2} \sigma_{i}^2(\Delta_1,\Delta_2). \nonumber
\ee
We find that for both the cell model and power spectrum model the largest term in $\hat{\sigma}_{i}^2$ when $\delta \Delta = 0$ is proportional to $1/\delta \Delta$ when 
$\delta \Delta \gg 1$.  If $\delta \Delta \gg 1$, then $\bar{\sigma}_{i}^2 \ll \Pbar^2$.  Similarly the variance of $\Pphi$ is also much smaller than $\Pbar^2$ when $\delta \Delta
\gg 1$ so that, to a good approximation, we have $\Pphi = \Pbar$.

\end{document}